\documentclass[lettersize,journal]{IEEEtran}
\usepackage{amsmath,amsfonts}
\DeclareMathOperator{\logit}{logit}
\newcommand{\AttnV}{\mathrm{Attn}_{v}}
\usepackage{algorithmic}
\usepackage{algorithm}
\usepackage{array}
\usepackage[caption=false,font=normalsize,labelfont=sf,textfont=sf]{subfig}
\usepackage{textcomp}
\usepackage{stfloats}
\usepackage{url}
\usepackage{verbatim}
\usepackage{cite}
\usepackage{graphicx}
\usepackage{booktabs}
\usepackage{multirow}
\usepackage[table]{xcolor}
\definecolor{lightblue}{RGB}{230, 242, 255}
\usepackage{threeparttable}
\usepackage{amssymb}
\usepackage{makecell}
\usepackage{placeins}

\begin{document}

\title{M2S-AVSR: Modality-aware Multi-view Self-supervised Representation for Robust Audio-Visual Speech Recognition}

\author{Fei~Su,
        Cancan~Li,
        Ming~Li,~\IEEEmembership{Senior Member, IEEE},
        and Juan~Liu,~\IEEEmembership{Senior Member, IEEE}

\thanks{
Fei Su, Cancan Li, and Juan Liu are with the School of Artificial Intelligence and the School of Computer Science, Wuhan University, China
(e-mail: \{fei.su, cancan.li, liujuan\}@whu.edu.cn).
}
\thanks{
Ming Li is with the School of Artificial Intelligence, The Chinese University of Hong Kong, Shenzhen, China and the School of Artificial Intelligence, Wuhan University, China
(e-mail: mingli369@cuhk.edu.cn).
}
\thanks{
Corresponding authors: Juan Liu and Ming Li.
}
}


\maketitle
\begin{abstract}
Audio-Visual Speech Recognition (AVSR) enhances speech recognition robustness by leveraging visual cues, while real-world scenarios remain challenging due to viewpoint variation, audio distortion, and visual occlusion, which degrade modality quality and increase audio-visual asynchrony. In this paper, we propose a novel Modality-aware Multi-view Self-supervised representation framework for robust Audio-Visual Speech Recognition (M2S-AVSR). First, we introduce a multi-view representation learning encoder to learn view-invariant visual speech representations. Next, we employ a modality-aware module that explicitly models modality quality and cross-modal synchrony to perform fine-grained modality-aware fusion, enabling fine-grained visual information injection during decoding. In addition, we release AISHELL8-RealScene, a public multi-scenario, multi-view conversational audio-visual dataset recorded in real-world environments, and establish a speech recognition benchmark on it. Experiments on English and Mandarin benchmarks demonstrate the effectiveness of the proposed method under challenging conditions. On LRS3, M2S-AVSR achieves up to 29.4\% relative improvement under viewpoint perturbation and visual degradation settings. Our method also achieves new state-of-the-art performance on the MISP2021-AVSR test set. On AISHELL8-RealScene, it achieves the best result in outdoor scenes. The proposed method and dataset provide useful support for future research on robust speech and multimodal tasks under realistic conditions.
\end{abstract}


\section{Introduction}
\label{sec:intro}


\IEEEPARstart{S}{poken} language is a major form of human communication, and Automatic Speech Recognition (ASR) is an important technology for human--machine interaction. In recent years, deep learning has significantly advanced ASR systems~\cite{graves2013speech,kim2017joint,parcollet2020e2e}, leading to strong performance under controlled conditions. Despite these advances, robust speech recognition in real-world environments remains challenging. In practical scenarios, microphone signals are often degraded by background noise, reverberation, competing speakers, and other forms of interference, which can substantially reduce recognition accuracy~\cite{wang2020complex,braun2021towards,tan2019learning}. To address these limitations, recent studies~\cite{petridis2018end,hong2022visual,vaswani2017attention,gulati2020conformer} have explored the use of visual information, such as lip movements, to provide complementary cues when the acoustic signal is unreliable.

Audio-visual speech recognition (AVSR) has been widely studied to improve robustness under adverse acoustic conditions by jointly modeling audio and visual modalities. Early approaches explored supervised architectures such as Connectionist Temporal Classification (CTC) and sequence-to-sequence frameworks~\cite{graves2006ctc,afouras2018deep}, demonstrating that visual cues, especially lip movements, can provide complementary information when audio signals are degraded. With the development of multimodal learning, AVSR systems~\cite{petridis2018end,hong2022visual,zhang2022learning} have achieved substantial improvements over audio-only counterparts, particularly in noisy environments.

Recent progress in AVSR has been strongly driven by large-scale pretraining. Self-supervised learning (SSL) methods, such as wav2vec~\cite{baevski2020wav2vec}, WavLM~\cite{chen2022wavlm}, and Whisper~\cite{radford2023robust}, have substantially improved acoustic modeling. Extending this paradigm to multimodal settings, AV-HuBERT~\cite{shi2022avhubert} and related approaches learn robust visual speech representations from large-scale unlabeled audio-visual data, followed by fine-tuning on limited labeled datasets. To alleviate the scarcity of large-scale video data, recent studies explore adapting large-scale audio-pretrained models to audio-visual tasks~\cite{pan2022leveraging,simic2024self}, achieving competitive performance with reduced video data requirements. More recently, large language models (LLMs) have also been introduced into AVSR. Models based on Whisper~\cite{radford2023robust} and LLaMA variants have been combined with visual features for multimodal reasoning and decoding~\cite{rouditchenko2024whisper,yeo2024visual,Cappellazzo2025Large,yeo2025mms}. However, most existing AVSR systems are still developed under constrained conditions with stable frontal views~\cite{afouras2018lrs3}. In real scenes, audio and visual signals are frequently corrupted by multiple factors. Audio suffers from noise and distortion~\cite{wu2026purification}, while visual inputs vary with viewpoint, occlusion, motion blur, and partial missing observations~\cite{nagrani2021attention,xuan2020cross}. In addition, temporal mismatch between audio and visual streams further complicates multimodal modeling. 

\begin{figure}[t]
    \centering
    \includegraphics[width=1\columnwidth]{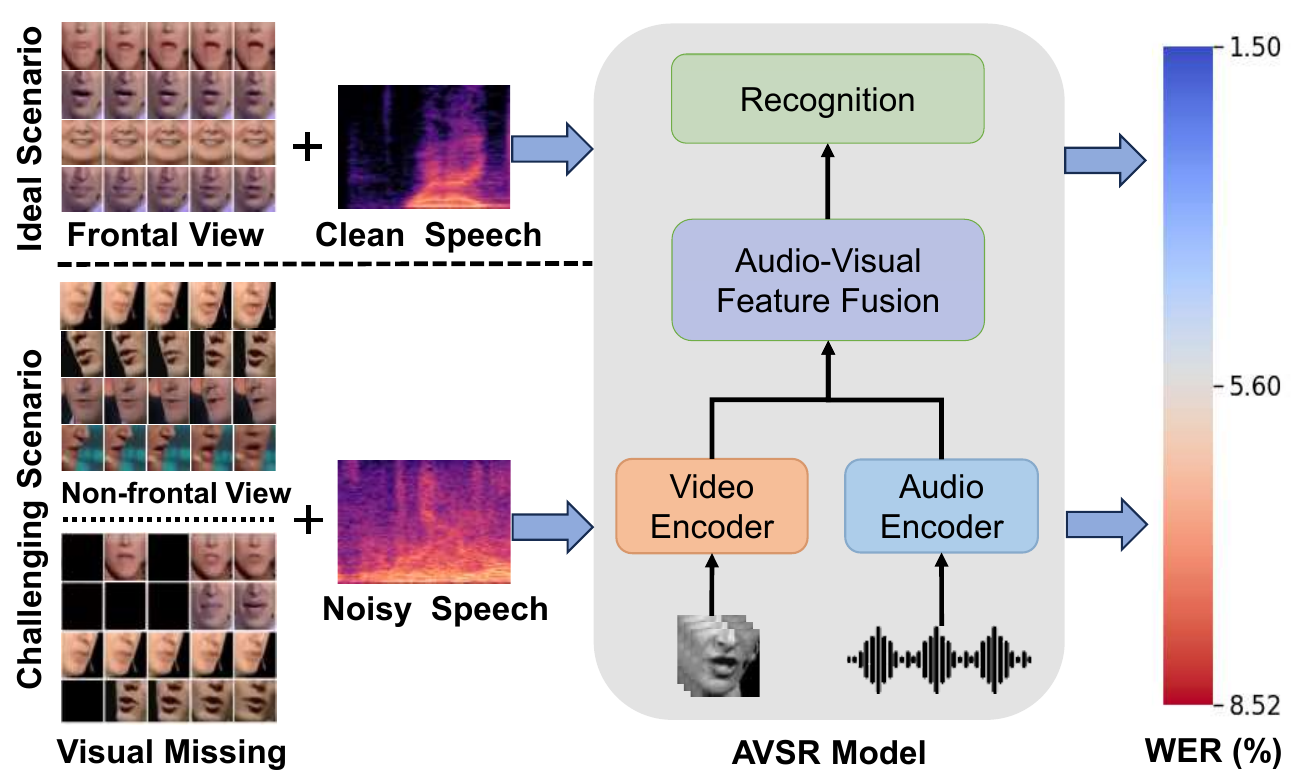}
        \caption{
        Illustration of AVSR performance degradation under challenging acoustic and visual conditions, evaluated on the LRS3 test set using Whisper-Flamingo~\cite{rouditchenko2024whisper} trained on 433\,h of labeled LRS3 data.
        }
    \label{fig_examples}
\end{figure}

As illustrated in Fig.~\ref{fig_examples}, we show an example of AVSR performance degradation using the Whisper-Flamingo~\cite{rouditchenko2024whisper} on the LRS3 test set~\cite{afouras2018lrs3}. Starting from clean frontal inputs, recognition accuracy degrades progressively with acoustic noise (SNR$=0~\mathrm{dB}$), viewpoint variation ($15^\circ$), and visual masking (ratio $=0.3$). This example shows that performance is sensitive to variations in visual conditions and their interaction with the audio stream. It indicates that AVSR systems~\cite{zhou2019modality,zhang2019robust,parthasarathy2020training} require improved modeling of multi-view variability and fine-grained audio-visual fusion.

To address these challenges, we propose \textbf{M2S-AVSR}, a Modality-aware Multi-view Self-supervised representation framework for robust Audio-Visual Speech Recognition in real-world environments. The proposed framework is an extension of our prior conference paper~\cite{su2025enhanced}. It is designed to improve AVSR robustness against viewpoint variation, modality degradation, and cross-modal inconsistency that commonly arise in practical scenarios. In particular, the framework combines robust multi-view visual representation learning with modality-aware audio-visual fusion, so that visual cues can be exploited more effectively under adverse conditions.

In addition, we release \textbf{AISHELL8-RealScene}\footnote{
The AISHELL8-RealScene dataset is publicly available at:
\url{https://huggingface.co/datasets/SMIIP-lab/AISHELL8-RealScene}.
The dataset is released under the CC BY-NC-SA 4.0 license.
}, a multi-scenario and multi-view audio-visual dataset collected in real-world environments. Compared with existing datasets, it places greater emphasis on diverse recording conditions and multi-view capture, providing a more realistic benchmark for studying robustness and generalization in AVSR.

Our main contributions are summarized as follows:

\begin{itemize}
    \item We propose a multi-view self-supervised visual representation learning strategy. By leveraging both real and synthesized views, the proposed method learns more robust visual speech representations across viewpoints.

    \item We propose a modality-aware fusion mechanism that explicitly considers modality quality and audio-visual synchrony during decoding, enabling more reliable multimodal integration under adverse acoustic and visual conditions.

    \item We release \textbf{AISHELL8-RealScene}, a public multi-scenario, multi-view conversational audio-visual dataset recorded in real-world environments, and establish a new benchmark for audio-visual speech recognition under realistic conditions.

    \item We conduct extensive experiments on LRS3, MISP2021-AVSR and AISHELL8-RealScene under challenging conditions, demonstrating the effectiveness and robustness of the proposed method.
\end{itemize}

\section{Related Works}
\label{sec:related}

\subsection{Audio-Visual Speech Recognition}

Audio-visual speech recognition (AVSR) integrates audio and visual modalities to improve robustness under adverse acoustic conditions. Early AVSR systems showed that visual cues, such as lip movements, can compensate for degraded audio signals~\cite{huang2013audio,mroueh2015deep}. With the development of deep learning, end-to-end AVSR models have achieved significant progress by jointly modeling multimodal inputs~\cite{ma2021conformer,hong2022visual}. The availability of large-scale datasets, such as LRS2 and LRS3~\cite{son2017lip,afouras2018lrs3}, together with Transformer-based architectures and improved fusion strategies~\cite{vaswani2017attention,gulati2020conformer}, has further advanced multimodal representation learning.

Self-supervised learning has become a key technique for AVSR. Methods such as AV-HuBERT~\cite{shi2022avhubert} learn joint audio-visual representations from large-scale unlabeled data and achieve strong performance after fine-tuning. Building upon this paradigm, subsequent studies extend pretrained speech models such as Whisper~\cite{radford2023robust} to multimodal AVSR~\cite{rouditchenko2024whisper}, while recent LLM-based approaches further improve contextual modeling and robust decoding ~\cite{yeo2024visual,Cappellazzo2025Large,yeo2025zero,anand2026mitigating}.

Despite their promising performance, these approaches mainly focus on improving recognition accuracy under relatively constrained conditions, often with high computational cost. Robustness under viewpoint variation, modality degradation, and cross-modal inconsistency in real-world environments remains insufficiently explored. To address this issue, we propose a unified framework for robust AVSR by combining multi-view self-supervised representation learning with modality-aware fusion.

\begin{figure*}[htbp] 
    \centering 
    \includegraphics[width=\textwidth]{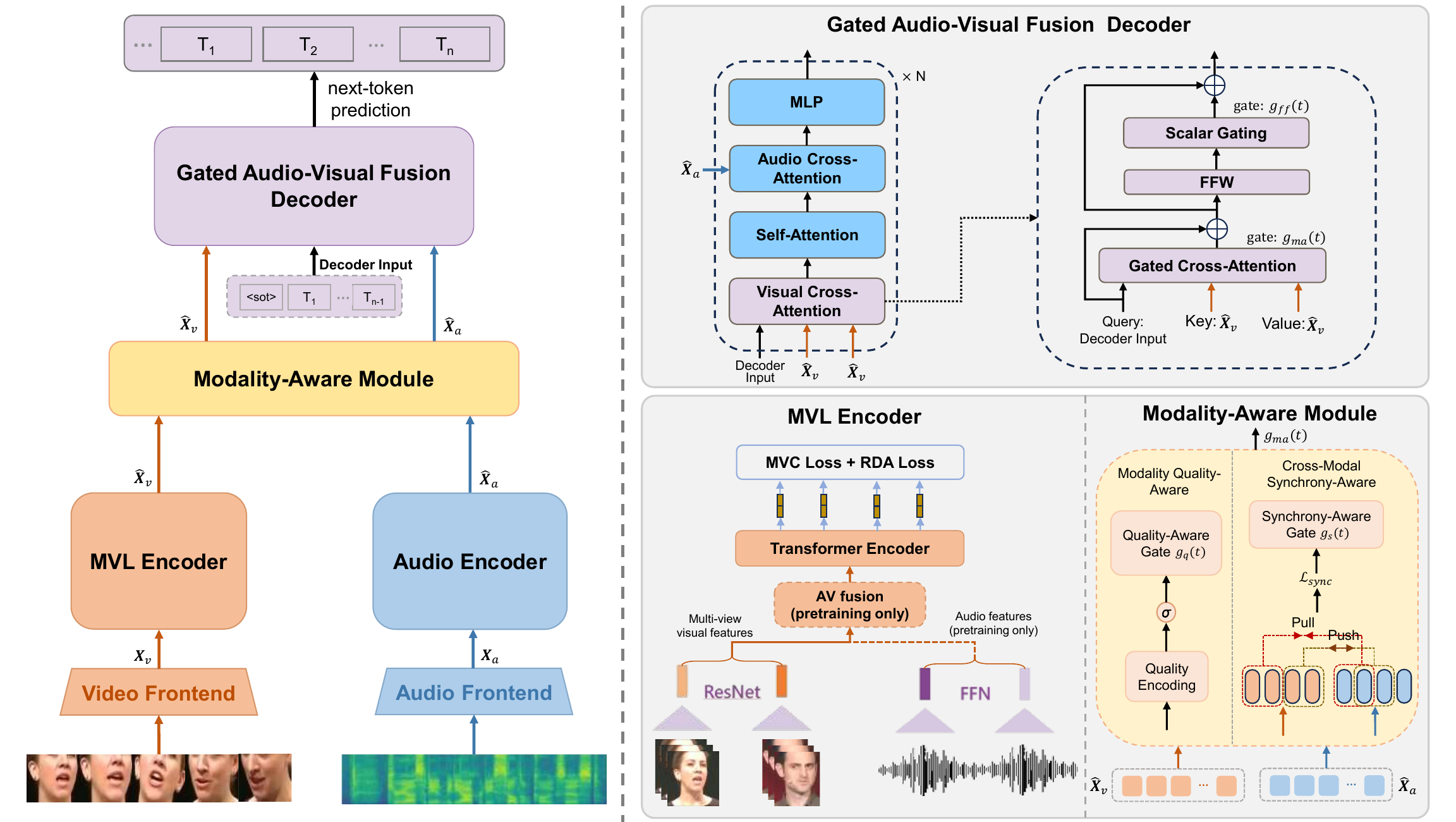}
    \caption{Overview of the proposed M2S-AVSR framework. \textbf{Left:} the overall architecture of the proposed framework. The audio and video front-ends use a two-layer Conv1D network and ResNet-18, respectively. The audio encoder and MVL encoder are initialized from Whisper and AV-HuBERT, respectively, and are built upon Transformer encoder architectures. \textbf{Right:} the three main components of M2S-AVSR, including the Gated Audio-Visual Fusion Decoder, the MVL Encoder, and the Modality-Aware Module. Dashed lines in the MVL encoder denote components used only during self-supervised pretraining. During AVSR training and inference, the MVL encoder takes only multi-view visual features, while the audio branch is disabled. MVC and RDA denote Multi-View Consistency and Representation Domain Alignment, respectively.}
    
    \label{fig_m2s} 
\end{figure*}

\subsection{Multi-View Representation Learning}

Learning representations that are robust to viewpoint changes is critical for visual speech modeling. In AVSR, most existing approaches focus on multimodal fusion rather than explicitly addressing viewpoint variability, which limits performance under non-frontal views or large pose variations~\cite{ma2021conformer}. 

Self-supervised learning has been widely explored to improve visual representation quality. Methods such as AV-HuBERT~\cite{shi2022avhubert} learn audio-visual representations from large-scale unlabeled data and achieve strong performance after fine-tuning. Many approaches~\cite{isobe2021multi,isobe2022efficient,axyonov2024audio} enforce consistency across different views or augmentations, enabling the learning of view-invariant visual patterns. Beyond pairwise objectives, recent works exploit multiple observations jointly to capture complementary information across views and enhance robustness through cross-view aggregation~\cite{zhao2022multi,zhu2023umiformer,sun2024vsformer}. Despite recent progress, multi-view representation learning still faces several challenges in real-world scenarios. Limited viewpoint diversity in training data restricts generalization to unseen poses or large viewing angles~\cite{das2023viewclr}. Moreover, many methods assume stable and complete observations, which is inconsistent with practical conditions where views may be missing, occluded, or degraded. 

To address these issues, we propose a multi-view self-supervised representation learning method that leverages both real and synthesized views to increase viewpoint diversity and improve robustness under missing or degraded visual conditions.

\subsection{Cross-modal Fusion}

Cross-modal fusion strategy determines how audio and visual information are combined to improve robustness and accuracy~\cite{huang2013audio}. Existing fusion strategies are commonly divided into early, middle, and late fusion according to the stage of integration. Representative implementations include feature-level concatenation~\cite{ma2021conformer,pan2022leveraging}, attention-based fusion~\cite{hong2022visual,simic2024self}, gated visual injection~\cite{rouditchenko2024whisper,yang2025injecting}, and other cross-modal transfer strategies~\cite{ZHANG2025111432,lim2025improving}. These strategies differ in the stage where modalities are integrated, leading to different trade-offs between cross-modal interaction and modality-specific robustness~\cite{baltruvsaitis2018multimodal,petridis2018audio,afouras2018deep}.

Recent studies further explore cross-modal supervision to improve representation alignment across modalities~\cite{nagrani2020disentangled,shi2022learning}. Related fusion strategies have also been studied in multimodal pretraining and vision-language modeling~\cite{sun2019videobert,radford2021learning,alayrac2022flamingo}. Nevertheless, most existing fusion methods rely on simple fusion strategies and lack comprehensive modeling of modality information under varying conditions. To address this limitation, we propose a modality-aware fusion mechanism that adaptively regulates cross-modal interaction for robust audio-visual speech recognition.

\section{Methods}
\label{sec:methods}

\subsection{Overall Architecture}
The proposed M2S-AVSR framework improves robustness in real-world audio-visual speech recognition through modality-aware multi-view self-supervised learning. As illustrated in Fig.~\ref{fig_m2s}, the M2S-AVSR framework adopts audio and visual front-ends to extract high-dimensional representations from the input speech waveform and lip video frames, respectively. Inspired by prior work on integrating visual features into Whisper~\cite{rouditchenko2024whisper}, we adopt a pretrained Whisper encoder as the audio front-end and initialize the visual front-end with AV-HuBERT to capture the lip motion representations. Unlike previous AVSR approaches that often lack robustness to viewpoint variations and perform audio-visual fusion in a uniform manner, the proposed framework introduces dedicated mechanisms to model both multi-view variability and fine-grained modality-aware factors for audio-visual speech recognition. Specifically, the \textbf{MVL encoder} is designed to learn robust visual speech representations across different views through multi-view self-supervised representation learning. Meanwhile, the \textbf{Modality-Aware Module} adaptively regulates the integration of visual information based on modality reliability and cross-modal consistency, enabling more robust audio-visual fusion under realistic conditions. 

\subsection{Multi-View Representation Learning Encoder}
\label{sec:mvl_encoder}

Our model adopts a pretrained Whisper encoder as the audio front-end and a multi-view representation learning encoder as the visual front-end. Specifically, given an input speech waveform, the Whisper encoder extracts audio representations $\mathbf{X}_a \in \mathbb{R}^{T_a \times D_a}$, where $T_a$ is the number of audio time steps and $D_a$ is the audio feature dimension. Meanwhile, lip image sequences are processed by the visual front-end to produce visual representations $\mathbf{X}_v \in \mathbb{R}^{T_v \times D_v}$, where $T_v$ and $D_v$ denote the temporal length and feature dimension of the visual representations, respectively.

Instead of directly using a pretrained AV-HuBERT as the visual encoder, we adopt a Multi-View representation Learning (MVL) encoder. It is initialized from an AV-HuBERT large model and further optimized for robustness to viewpoint variations. Following~\cite{su2025enhanced}, multi-view simulated visual data are generated and combined with real data to train the MVL encoder. A self-supervised strategy based on AV-HuBERT~\cite{shi2022avhubert} is employed to learn viewpoint-invariant and domain-aligned visual representations from these real-simulated pairs.

\subsubsection{Multi-View Consistency (MVC) Loss}

Inspired by self-supervised training for visual speech representation learning, we impose a consistency constraint between real samples and their synthesized multi-view counterparts to encourage viewpoint-invariant representations. Let $\{\mathbf{x}_{r,i}\}_{i=1}^{N}$ denote real visual samples, and let $\{\mathbf{x}_{s,i}\}_{i=1}^{N}$ denote the corresponding synthesized samples generated from the same utterances using the multi-view simulation strategy in \cite{su2025enhanced}. Let $f(\cdot)$ be the embedding function that maps these inputs into a latent space invariant to viewpoint changes. The corresponding visual embeddings are denoted as $\mathbf{X}_{r,i}=f(\mathbf{x}_{r,i})$ and $\mathbf{X}_{s,i}=f(\mathbf{x}_{s,i})$, respectively, where $\mathbf{X}_{r,i}, \mathbf{X}_{s,i} \in \mathbb{R}^{T_v \times D_v}$.

We first introduce an element-wise alignment term to reduce direct feature discrepancy between the real and synthesized views:
\begin{equation}
L_{\mathrm{mse}} =
\frac{1}{N}\sum_{i=1}^{N}
\left\|
\mathbf{X}_{r,i} - \mathbf{X}_{s,i}
\right\|_2^2 .
\end{equation}

However, feature-level similarity is insufficient to ensure consistent structural relationships within the learned representations. Therefore, we further introduce a correlation alignment term. Let $\mathbf{Z}_{r,i}, \mathbf{Z}_{s,i} \in \mathbb{R}^{T_v \times D_v}$ denote the normalized versions of $\mathbf{X}_{r,i}$ and $\mathbf{X}_{s,i}$. Their correlation discrepancy is defined as
\begin{equation}
L_{\mathrm{corr}} =
\frac{1}{N}\sum_{i=1}^{N}
\left\|
\mathrm{Corr}(\mathbf{Z}_{r,i}) -
\mathrm{Corr}(\mathbf{Z}_{s,i})
\right\|_F^2 .
\end{equation}
where $\mathrm{Corr}(\cdot)$ computes the correlation matrix and $\|\cdot\|_F$ denotes the Frobenius norm. This term enforces the real and synthesized views to share similar structural dependencies among latent dimensions, thereby improving semantic consistency under viewpoint changes.

The final multi-view consistency objective is defined as
\begin{equation}
L_{\mathrm{MVC}} =
\alpha L_{\mathrm{corr}} +
(1-\alpha)L_{\mathrm{mse}},
\end{equation}
where $\alpha \in [0,1]$ controls the trade-off between structural consistency and feature-level alignment.

\subsubsection{Representation Domain Alignment (RDA) Loss}
\label{sec:rda_loss}
To reduce the gap between real and simulated visual representations, we employ a contrastive objective to learn domain-invariant visual speech representations. Specifically, for each real sample $\mathbf{x}_{r,i}$, we select the corresponding synthesized sample $\mathbf{x}_{s,i}$ generated from the same utterance with minimal viewpoint deviation as a positive pair. Negative samples are constructed from other utterances within the same mini-batch. This objective encourages the encoder to focus on speech-related motion patterns while suppressing domain-specific artifacts introduced by synthesized samples.

Formally, let $f(\cdot)$ be the embedding function, the visual embeddings are defined as $\mathbf{X}_{r,i}=f(\mathbf{x}_{r,i})$ and $\mathbf{X}_{s,i}=f(\mathbf{x}_{s,i})$, where $\mathbf{X}_{r,i}, \mathbf{X}_{s,i} \in \mathbb{R}^{T_v \times D_v}$. The similarity between two samples is defined using cosine similarity:

\begin{equation}
\mathrm{sim}(\mathbf{X}_{r,i},\mathbf{X}_{s,i}) =
\frac{\left\langle \mathbf{X}_{r,i}, \mathbf{X}_{s,i} \right\rangle}
{\left\|\mathbf{X}_{r,i}\right\|\left\|\mathbf{X}_{s,i}\right\|}.
\end{equation}

The representation domain alignment loss is formulated as:
\begin{equation}
L_{\mathrm{RDA}} =
-\frac{1}{N}\sum_{i=1}^{N}
\log
\frac{\exp(\mathrm{sim}(\mathbf{X}_{r,i},\mathbf{X}_{s,i})/\tau)}
{\sum_{j=1}^{M}\exp(\mathrm{sim}(\mathbf{X}_{r,i},\mathbf{X}_{n,j})/\tau)},
\end{equation}

where $N$ is the number of real--synthetic positive pairs, $M$ is the number of negative samples, $\tau$ is a temperature parameter, and $\mathbf{X}_{n,j}$ denotes the embedding of the $j$-th negative sample.

Finally, the MVL encoder is trained by jointly optimizing the multi-view consistency loss $L_{\mathrm{MVC}}$, the representation domain alignment loss $L_{\mathrm{RDA}}$, and the masked prediction objective $L_{\mathrm{MMP}}$, which corresponds to the masked multimodal cluster prediction loss used in AV-HuBERT pretraining~\cite{shi2022learning}. The MVL objective is defined as:

\begin{equation}
L_{\mathrm{MVL}} =
\lambda_{\mathrm{MVC}} L_{\mathrm{MVC}}
+
\lambda_{\mathrm{RDA}} L_{\mathrm{RDA}}
+
\lambda_{\mathrm{MMP}} L_{\mathrm{MMP}},
\end{equation}

where $L_{\mathrm{MVL}}$ denotes the overall loss for training the MVL encoder, $\lambda_{\mathrm{MVC}}$, $\lambda_{\mathrm{RDA}}$, and $\lambda_{\mathrm{MMP}}$ are weighting factors. The learned representations are denoted as $\hat{\mathbf{X}}_v$ and are used as the visual input of the Modality-Aware Module.

\begin{figure}[t]
    \centering
    \includegraphics[width=1\columnwidth]{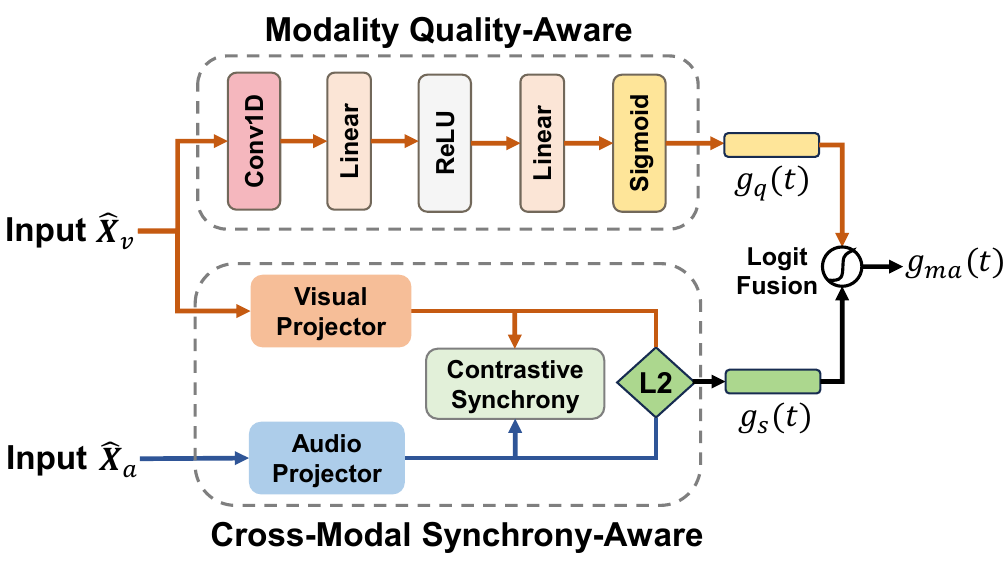}
        \caption{Detailed structure of the modality-aware module. The upper branch estimates the visual quality score $g_q(t)$. The lower branch estimates the synchrony score $g_s(t)$ through contrastive synchrony modeling and local-window $\ell_2$ distance. The two scores are then fused to produce the final modality-aware gating value $g_{\mathrm{ma}}(t)$.}
    \label{fig_ma_module}
\end{figure}

\subsection{Modality-Aware Modeling}
	\label{sec:modality_aware}
	
Based on the learned multi-view visual representation $\hat{\mathbf{X}}_v$, we further consider how visual information should be integrated with the audio stream under realistic conditions. Since audio and visual modalities may be degraded by noise, occlusion, motion blur, and temporal misalignment, directly performing uniform multimodal fusion may introduce unreliable visual cues into the decoder. To alleviate this issue, we retain the Whisper-based audio representation as a stable acoustic foundation and introduce a modality-aware mechanism to regulate visual information injection according to modality quality and cross-modal synchrony.

As shown in Fig.~\ref{fig_ma_module}, let $\hat{\mathbf{X}}_a \in \mathbb{R}^{B \times T_a \times D_a}$ and $\hat{\mathbf{X}}_v \in \mathbb{R}^{B \times T_v \times D_v}$ denote the audio and visual representations extracted by the Whisper encoder and the MVL encoder, respectively.

\subsubsection{Modality Quality-Aware}
\label{sec:quality_awareness}

Visual modality reliability may vary significantly under realistic conditions due to occlusion, motion blur, and viewpoint changes. As a result, uniformly injecting visual features into the decoder may introduce unreliable visual cues and degrade recognition performance. To address this issue, we estimate a frame-level quality-aware gate from the visual representation to characterize its time-varying reliability during decoding.

We first apply a temporal convolution to $\hat{\mathbf{X}}_v$ to capture local contextual patterns:
\begin{equation}
\mathbf{C}_v = \mathrm{Conv1D}(\hat{\mathbf{X}}_v) \in \mathbb{R}^{B \times T_v \times D_v}.
\end{equation}

Then, following a lightweight gating design, an intermediate representation is computed as
\begin{equation}
\mathbf{H}_q(t)=\mathrm{ReLU}\!\left(\mathbf{C}_v(t)\mathbf{W}_{q1}+\mathbf{b}_{q1}\right).
\end{equation}

The modality quality gate $g_q(t)\in[0,1]$ is then obtained from the intermediate representation through a linear transformation followed by a sigmoid activation:
\begin{equation}
g_q(t)=\sigma\!\left(\mathbf{H}_q(t)\mathbf{W}_{q2}+\mathbf{b}_{q2}\right),
\end{equation}
where $\sigma(\cdot)$ denotes the sigmoid function.

	
\subsubsection{Cross-Modal Synchrony-Aware}
\label{sec:synchrony_awareness}

High-quality visual representations do not necessarily imply reliable cross-modal fusion, particularly under noisy conditions where audio and visual streams may become locally inconsistent. In such cases, excessive reliance on visual cues may introduce unreliable modality bias, especially for acoustically ambiguous or visually similar speech units. To alleviate this issue, we further model audio--visual synchrony to dynamically regulate the contribution of visual information according to the consistency between the two modalities.

Specifically, the audio and visual representations are projected into a shared synchrony embedding space~\cite{chung2016out,nagrani2020disentangled,he2022end} through modality-specific projection networks, yielding $\mathbf{E}_a = \mathrm{Proj}_a(\hat{\mathbf{X}}_a)$ and $\mathbf{E}_v = \mathrm{Proj}_v(\hat{\mathbf{X}}_v)$, where $\mathrm{Proj}_a(\cdot)$ and $\mathrm{Proj}_v(\cdot)$ denote the audio and visual synchrony projection networks, respectively. The visual representations are temporally aligned with the audio representations within the visual projector. The two projectors then process the visual and audio features using lightweight temporal convolution followed by point-wise projection, mapping the two modalities into a comparable embedding space.

The learned embeddings capture cross-modal temporal consistency, where synchronized audio--visual representations yield smaller distances than inconsistent ones. Based on these embeddings, we estimate the synchrony between modalities using a local-window average $\ell_2$ distance:
\begin{equation}
    D_{\mathrm{s}}(t)
    =
    \frac{1}{2T_w+1}
    \sum_{k=t-T_w}^{t+T_w}
    \left\|
    \mathbf{E}_a(k)-\mathbf{E}_v(k)
    \right\|_2,
\end{equation}
where $\mathbf{E}_a(k), \mathbf{E}_v(k) \in \mathbb{R}^{D_s}$ denote the audio and visual synchrony embeddings at time step $k$, respectively. Here, $T_w$ defines a small temporal window that provides local tolerance to slight audio--visual misalignment. The distance is converted into a synchrony-aware gate:
\begin{equation}
   g_s(t)
    =
    \frac{\gamma}{\gamma+D_{\mathrm{s}}(t)},
\end{equation}
where $\gamma>0$ is a scaling constant. The gate $g_s(t)$ is used for subsequent audio--visual feature fusion. In our experiments, we set $\gamma=1$, which normalizes $g_s(t)$ to the range $(0,1]$.

\subsubsection{Modality-Aware Fusion}
\label{sec:modality_fusion}

The quality-aware gate $g_q(t)$ and the synchrony-aware gate $g_s(t)$ are further integrated into a unified modality-aware gate $g_{\mathrm{ma}}(t)$ to jointly model visual reliability and cross-modal consistency during decoding. This formulation enables adaptive visual information injection according to both visual reliability and cross-modal consistency. To place the two gates on a shared score scale, the logit function $\logit(p)=\log\frac{p}{1-p}$ is adopted~\cite{satopaa2014combining}. This transformation ensures that positive values increase the fused gating score, while negative values decrease it. The modality-aware gate is defined as~\cite{fan2024giant}:
\begin{equation}
    g_{\mathrm{ma}}(t)
    =
    \tanh\!\Big(
    w_q\,\logit(g_q(t)) + w_s\,\logit(g_s(t))
    \Big),
\end{equation}
where $w_q$ and $w_s$ are learnable scalars. They are initialized with small magnitudes and refined during training.

\begin{figure*}[t] 
    \centering 
    \includegraphics[width=\textwidth]{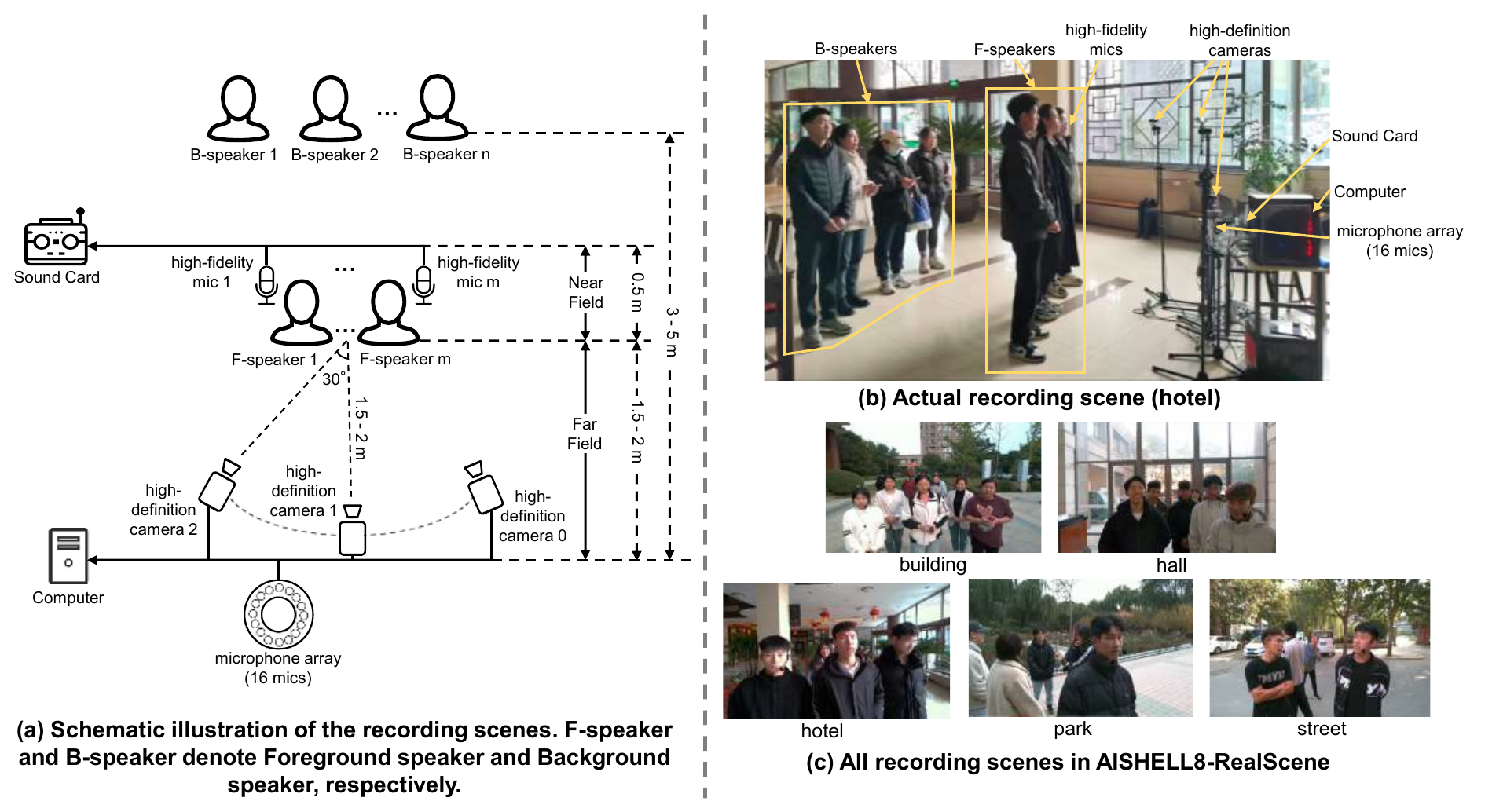}
    \caption{Overview of the AISHELL8-RealScene recording scenes. (a) Schematic illustration of the recording scenes, where $m$ foreground speakers (F-speakers, $m\!=\!1\text{--}5$) converse in Mandarin, while $n$ background speakers (B-speakers, $n\!=\!1\text{--}5$) stand behind them to create natural interfering activities. (b) An example of an actual recording scene (hotel). (c) All recording scenes in AISHELL8-RealScene: outside the office \textbf{building}, residential \textbf{hall}, \textbf{hotel}, \textbf{park}, and \textbf{street}.}
    \label{fig_aishell8rs_overview} 
\end{figure*}

\subsection{Gated Audio-Visual Fusion Decoder}
\label{sec:gated_decoder}

Directly introducing visual features into a pretrained Whisper decoder may disturb the originally learned acoustic and linguistic representations, particularly when the visual modality becomes unreliable under varying conditions. To achieve more stable audio--visual fusion, we integrate the proposed modality-aware gating mechanism into a Whisper-based decoder~\cite{alayrac2022flamingo,rouditchenko2024whisper}, where a gated visual cross-attention layer is inserted at the beginning of each decoder block to adaptively regulate visual information injection. 

Specifically, the decoder hidden state is used as the query, and the visual representation from the MVL encoder is used as the key and value. The gated cross-attention layer is defined as follows, where $\mathbf{h}$ is the input to the decoder block, $\hat{\mathbf{X}}_v$ are the visual features, $\AttnV$ is multi-head cross-attention, and $\mathrm{LN}(\cdot)$ denotes LayerNorm:
\begin{equation}
	\mathbf{h}_{\mathrm{attn}}
	=
	\mathbf{h}
	+
	g_{\mathrm{ma}}(t)\cdot
	\AttnV\!\big(\mathrm{LN}(\mathbf{h}), \hat{\mathbf{X}}_v\big).
\end{equation}

For the subsequent feed-forward branch, we keep a per-block learnable gate and apply
$\mathbf{h}_{\mathrm{out}}=\mathbf{h}_{\mathrm{attn}}+g_{\mathrm{ff}}\cdot \mathrm{FFW}\!\big(\mathrm{LN}(\mathbf{h}_{\mathrm{attn}})\big)$,
where $\mathrm{FFW}(\cdot)$ is an MLP and $g_{\mathrm{ff}}$ is a learnable scalar gate bounded by $\tanh(\cdot)$ for stable scaling. All gating parameters are initialized to yield near-zero gated updates from the beginning.

\subsection{Training Strategy}
\label{sec:training_strategy}

The proposed framework is trained in three stages. First, the MVL encoder is initialized from AV-HuBERT~\cite{shi2022learning} and pre-trained on multi-view visual data, as described in Section~\ref{sec:rda_loss}. Second, following~\cite{rouditchenko2024whisper}, all layers of the Whisper encoder are fine-tuned on audio-only data for domain adaptation. Third, with both the Whisper encoder and the MVL encoder frozen, the modality-aware module is optimized using a contrastive objective~\cite{chung2016out}. Temporally aligned audio--visual segments are treated as positive pairs, while temporally shifted audio segments paired with the same visual input are treated as negative pairs. This objective encourages synchronized embeddings to be close and misaligned ones to be separated. The synchrony loss is defined as:
\begin{equation}
L_{\mathrm{sync}}
=
\frac{1}{T_s}\sum_{t=1}^{T_s}
\left(
y_t D_{\mathrm{s}}^2
+
(1-y_t)[\max(m-D_{\mathrm{s}},0)]^2
\right)
\end{equation}
where $T_s$ denotes the number of aligned time steps, $y_t\in\{0,1\}$ is a binary similarity indicator between the audio and visual inputs, and $m$ is a margin that controls the separation between aligned and misaligned pairs.

Given an input audio--visual pair, the decoder generates the target token sequence in an autoregressive manner conditioned on both audio and visual features. Let $p_{\mathrm{att}}(\mathbf{s}|\hat{\mathbf{X}}_a,\hat{\mathbf{X}}_v)$ denote the attention-based posterior probability. The attention-based objective is defined as~\cite{watanabe2017hybrid}:
\begin{equation}
	p_{\mathrm{att}}(\mathbf{s}|\hat{\mathbf{X}}_a,\hat{\mathbf{X}}_v)
	=
	\prod_{l=1}^{L}
	p(s_l|s_1,\ldots,s_{l-1},\hat{\mathbf{X}}_a,\hat{\mathbf{X}}_v),
\end{equation}
\begin{equation}
	L_{\mathrm{att}}
	=
	-\sum_{l=1}^{L}
	\log p(s_l|s_1,\ldots,s_{l-1},\hat{\mathbf{X}}_a,\hat{\mathbf{X}}_v),
\end{equation}
where $\mathbf{s}=\{s_1,\ldots,s_L\}$ denotes the target token sequence, and $\hat{\mathbf{X}}_a$ and $\hat{\mathbf{X}}_v$ represent the audio and visual representations extracted by the frozen encoders, respectively.

The training objective for audio-visual fusion is defined as
\begin{equation}
L_{\mathrm{avsr}}
=
L_{\mathrm{att}}
+
\lambda_{\mathrm{sync}} L_{\mathrm{sync}},
\end{equation}
where $L_{\mathrm{att}}$ is the primary attention-based recognition objective, and $L_{\mathrm{sync}}$ is an auxiliary synchrony regularization term. $\lambda_{\mathrm{sync}}$ is a weighting factor, which is set to 0.1 in our experiments.

\begin{figure}[t]
    \centering
    \includegraphics[width=1.03\columnwidth]{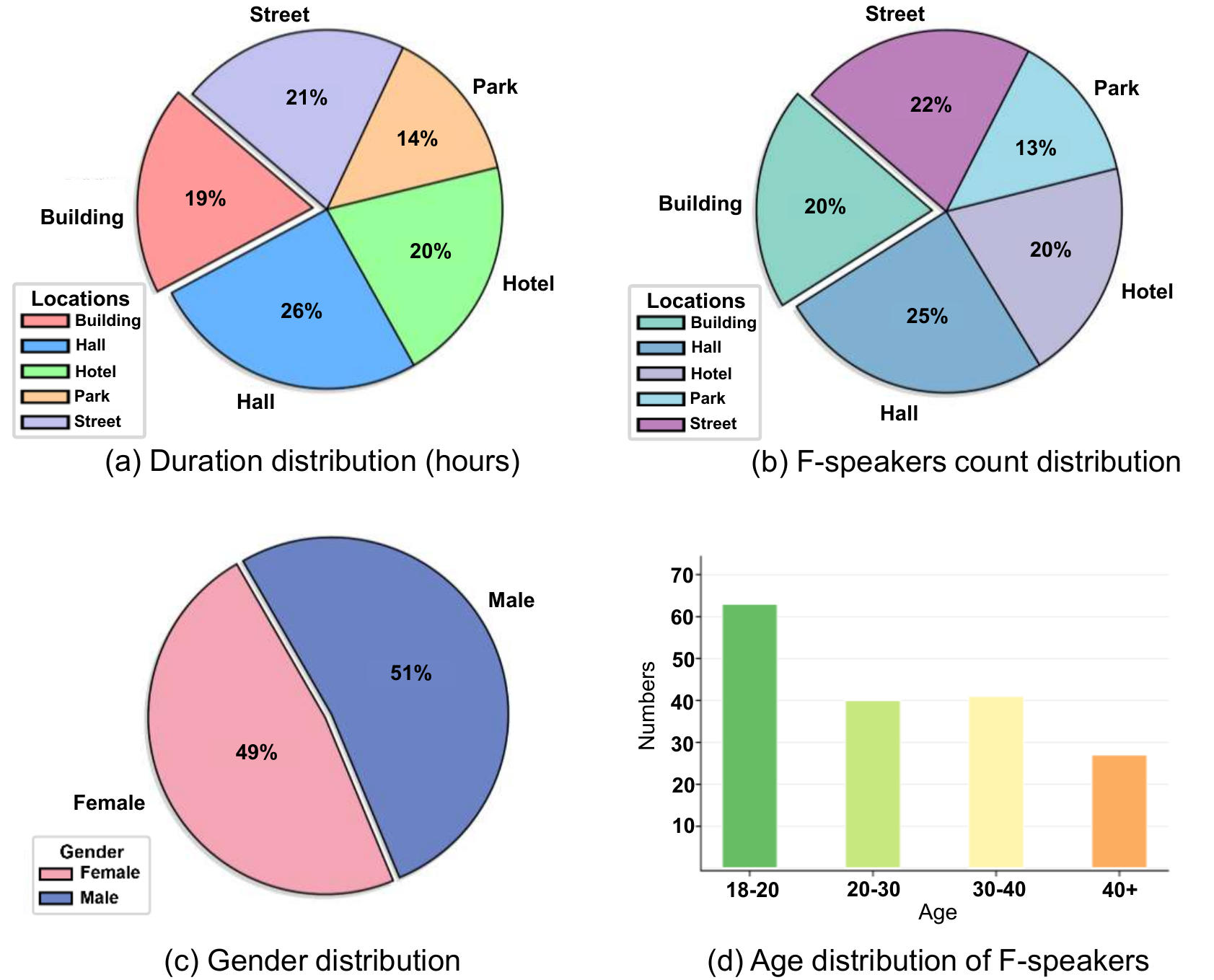}
        \caption{Distribution analysis of AISHELL8-RealScene dataset.}
    \label{fig_aishell8rs_statistics}
    \vspace{-10pt}
\end{figure}

\subsection{AISHELL8-RealScene Dataset}
\label{sec:aishell8rs}

\subsubsection{Dataset Description}
Regarding prior audio-visual datasets \cite{son2017lip,afouras2018lrs3,zhao2019cascade,chen2022audio,chen2023cn,Anwar2023MuAViCAM,zhou2025ave}, open-source real-world corpora are still scarce, especially those with outdoor scenes. This makes it difficult to train and evaluate models under diverse real-world conditions.
To address this issue, we collect and release \textbf{AISHELL8-RealScene}, a public multi-scenario, multi-view audio-visual speech corpus recorded in real-world scenes. It is designed to support research on speech and related tasks under realistic conditions and was collected by the database company AISHELL. The dataset contains 102.19 hours of synchronized audio and video from 171 speakers (foreground speakers).
 
 As illustrated in Fig.~\ref{fig_aishell8rs_overview}(a) and (b), AISHELL8-RealScene is collected with a synchronized multi-device setup in natural environments with real-world background noise. In each session, $m$ foreground speakers ($m=1\text{--}5$) either interact with an intelligent device or converse with other F-speakers, while $n$ background speakers ($n=1\text{--}5$) stay behind them and introduce interference such as background speech, phone calls, walking, and queuing. This design creates realistic acoustic and visual disturbances while addressing privacy concerns. Near-field audio is recorded only for the F-speakers, while multi-channel far-field audio is also provided. To reduce cross-device asynchrony, all recording devices are synchronized before recording, and a prompt alignment sound is used for manual alignment during post-processing.

As illustrated in Fig.~\ref{fig_aishell8rs_overview}(c), AISHELL8-RealScene contains recordings from five locations: outside the office \textbf{building}, residential \textbf{hall}, \textbf{hotel}, \textbf{park}, and \textbf{street}. The dialogue topics are location-specific and reflect natural daily interactions. To cover diverse conversational settings and realistic interference, we define five permutation configurations (P1--P5) by varying the numbers of foreground and background speakers. Here, ``interacts with an intelligent device'' means that the main F-speaker faces the recording devices and speaks as if conversing with an intelligent device.
\begin{itemize}
	\item \textit{Permutation 1 (P1)}: One F-speaker interacts with an intelligent device, while 1--2 B-speakers queue behind and generate interference.
	\item \textit{Permutation 2 (P2)}: Same as P1, but with 3--5 B-speakers.
	\item \textit{Permutation 3 (P3)}: One F-speaker interacts with an intelligent device, with 3--5 B-speakers. One or two B-speakers queue behind the F-speaker, while the remaining B-speakers only introduce interference.
	\item \textit{Permutation 4 (P4)}: Two F-speakers with 3--5 B-speakers, where the main F-speaker either interacts with the device or dialogues with the other F-speaker, while B-speakers follow the same protocol as in P3.
	\item \textit{Permutation 5 (P5)}: 3--5 F-speakers with 1--5 B-speakers, maintaining the same setting as in P4.
\end{itemize}

\begin{table}[!t]
	\centering
	\caption{Location and group configuration of AISHELL8-RealScene. ``Groups'' denotes the number of recorded conversation groups at each location, and each group follows one of the permutation settings (P1--P5).}
	\label{tabel:aishell8rs_loc_group}
	\scriptsize
	\setlength{\tabcolsep}{4.2pt}
	\renewcommand{\arraystretch}{1.12}
	\begin{tabular}{c l c c c c c c c c}
		\toprule
		\textbf{ID} & \textbf{Location} & \textbf{Indoor} & \textbf{Outdoor} & \textbf{Groups} & \textbf{P1} & \textbf{P2} & \textbf{P3} & \textbf{P4} & \textbf{P5} \\
		\midrule
		L1 & building &  & \checkmark & 14 & 1 & 2 & 1 & 4 & 6 \\
		L2 & hall     & \checkmark &  & 15 & 1 & 1 & 1 & 4 & 8 \\
		L3 & hotel    & \checkmark &  & 14 & 2 & 1 & 1 & 4 & 6 \\
		L4 & park     &  & \checkmark & 10 & 1 & 1 & 1 & 2 & 5 \\
		L5 & street   &  & \checkmark & 17 & 2 & 1 & 1 & 4 & 9 \\
		\bottomrule
	\end{tabular}
	\vspace{-2mm}
\end{table}

\begin{table}[t]
	\caption{Dataset split statistics of AISHELL8-RealScene. ``Sessions'' denotes the number of recording sessions.}
	\label{table:aishell8rs_data_split}
	\centering
	\footnotesize
	\renewcommand{\arraystretch}{1.12}
	\setlength{\tabcolsep}{7.0pt}
	\scalebox{1.0}{
		\begin{tabular}{c c c c @{\hspace{6pt}}|@{\hspace{6pt}} c}
			\toprule
			\textbf{Statistic} & \textbf{Train} & \textbf{Dev} & \textbf{Eval} & \textbf{Total} \\
			\midrule
			Duration (h)         & 79.84 & 10.70 & 11.65 & 102.19 \\
			Indoor duration (h)  & 36.74 & 5.03  & 5.28  & 47.05  \\
			Outdoor duration (h) & 43.10 & 5.67  & 6.37  & 55.14  \\
			Sessions             & 133   & 18    & 20    & 171    \\
			Groups               & 56    & 7     & 7     & 70     \\
			F-speakers           & 133   & 18    & 20    & 171    \\
			Gender ratio (M/F)   & 1:1.18 & 1:0.29 & 1:0.82 & 1:0.99 \\			
			\bottomrule
	\end{tabular}}
	\vspace{-2mm}
\end{table}

As shown in Table~\ref{tabel:aishell8rs_loc_group}, we recorded multiple conversation groups at each location, and the speakers in each group are configured according to the permutation settings (P1--P5).

\subsubsection{Data Recording and Processing}
\label{sec:aishell8rs_processing}
\textbf{Audio data.} Far-field audio is captured by a circular microphone array (16 microphones, 16\,kHz, 16-bit) facing the foreground speakers. To reduce data size while preserving spatial information, we release 8-channel audio by uniformly selecting one microphone every $45^\circ$ from the array. In addition, each foreground speaker wears a near field close-talking microphone for transcription. Each recording session contains multiple dialogues lasting 30--40 minutes. After processing, near-field audio is provided as single-channel 16\,kHz signals, and far-field audio as 8-channel 16\,kHz signals. Detailed device specifications are described in the dataset manual.

\textbf{Video data.} Video is captured by three HD cameras (RGB 1280$\times$720, 25\,fps) from multiple viewpoints. As shown in Fig.~\ref{fig_aishell8rs_overview}(a), the cameras are placed at different horizontal positions with an angular interval of approximately $30^\circ$, denoted as D0, D1, and D2 from left to right. To protect privacy and reduce irrelevant visual content, we apply face detection~\cite{guo2021sample}, face recognition~\cite{deng2019arcface}, and face re-identification~\cite{douze2024faiss,johnson2019billion} to extract only the foreground-speaker face region. The released face crops are provided in $256\times256$ resolution at 25\,fps.

\subsubsection{Data Distribution}
\label{sec:aishell8rs_distribution}
As shown in Fig.~\ref{fig_aishell8rs_statistics}, AISHELL8-RealScene is carefully designed with balanced data allocation and comprehensive speaker coverage to support a wide range of speech-related research under real-world conditions. 
The recording durations and F-speaker counts are distributed relatively evenly across the five locations (Fig.~\ref{fig_aishell8rs_statistics}(a) and (b)), ensuring that each location contributes a comparable amount of data rather than being dominated by a single scene. The speakers assigned to different locations are non-overlapping. Fig.~\ref{fig_aishell8rs_statistics}(c) and (d) show the gender and age distributions of the speakers, indicating a near-balanced gender composition and broad age coverage for downstream modeling.

As shown in Table~\ref{table:aishell8rs_data_split}, we split AISHELL8-RealScene into training, development, and evaluation sets, accounting for 79.84\%, 10.70\%, and 11.65\% of the data, respectively.
The ratio of overlapping-speech segments to all speech segments is about 25\%. No speakers are shared across the three subsets. The recording sessions are partitioned without overlap across subsets. All three subsets cover the five recording locations, ensuring that each split is both representative and comparable under consistent environments.

\section{Experimental Settings}
\label{sec:exp_settings}

\subsection{Datasets}
\label{sec:datasets}
The proposed M2S-AVSR framework extends our previous work~\cite{su2025enhanced} by introducing modality-aware fusion for improved robustness in real-world scenarios. We evaluate it on AISHELL8-RealScene, LRS3~\cite{afouras2018lrs3}, VoxCeleb2~\cite{Chung18b}, MISP2021-AVSR~\cite{chen2022first,chen2022audio}, and OuluVS2~\cite{Anina2015OuluVS2AM}. Following~\cite{shi2022learning}, LRS3 is split into 433\,h for training, 1\,h for validation, and 1\,h for testing. The LRS3 training set is further combined with 1,326 hours of English videos from VoxCeleb2, forming a 1,759\,h collection for large-scale pretraining. For noisy training, additive noise is applied following prior work~\cite{makino2019recurrent,ma2023auto,Anwar2023MuAViCAM,rouditchenko2024whisper}, where ``natural'', ``music'', and ``babble'' noises are sampled from MUSAN~\cite{snyder2015musan}, and overlapping speech noise is sampled from LRS3. MISP2021-AVSR is a Mandarin conversational AVSR dataset recorded in realistic home entertainment environments, containing 106.09 hours of training data. We follow~\cite{chen2022first} and evaluate on far-field audio and video. OuluVS2 provides real multi-view recordings from five fixed viewpoints (0$^\circ$, 30$^\circ$, 45$^\circ$, 60$^\circ$, and 90$^\circ$), enabling evaluation of robustness to real-world viewpoint variations.
\subsubsection{Audio Data Preprocess}
We use Whisper Large~\cite{radford2023robust} as the audio encoder. For each utterance, 80-dimensional log-Mel filterbank features are extracted from 16\,kHz audio using a 25\,ms window and a 10\,ms frame shift. For MISP2021-AVSR and AISHELL8-RealScene, far-field audio is further enhanced by weighted prediction error (WPE)~\cite{drude2018nara} and guided source separation (GSS)~\cite{boeddeker2018front}. During Whisper fine-tuning, we apply speed perturbation, SpecAug~\cite{park2019specaugment}, and continuous segment splicing, while only SpecAug is used during AVSR training.

\subsubsection{Video Data Preprocess}
The video preprocessing follows~\cite{rouditchenko2024whisper}. All video streams are sampled at 25 fps and converted to grayscale. The mouth region is cropped with a bounding box of size $96 \times 96$. During training, a random crop of size $88 \times 88$ and horizontal flipping with probability 0.5 are applied. During inference, a center crop of size $88 \times 88$ is used.

\subsubsection{Synthesized Data}
Since AISHELL8-RealScene already provides multi-view data, synthesized multi-view data are mainly constructed for LRS3 using the pipeline in~\cite{su2025enhanced}. We use 150 epochs for geometry offset optimization, 100 for texture refinement, and 80 for joint geometry-texture optimization. Virtual camera viewpoints are sampled from $-25^\circ$ to $25^\circ$ at $5^\circ$ intervals. Based on the synthesized data, we construct multi-view extended sets, where ms30h contains 30\% synthesized and 70\% real data, while ms433h and ms1759h contain 40\% synthesized and 60\% real data.

Recognition performance is evaluated using standard error-rate metrics. For the English dataset LRS3, we report the word error rate (WER), while for the Mandarin datasets MISP2021-AVSR and AISHELL8-RealScene, we report the character error rate (CER). Both metrics are computed using the Levenshtein distance between the predicted transcription and the reference text, namely $\mathrm{WER}/\mathrm{CER} = (S + D + I)/N$, where $S$, $D$, and $I$ denote the numbers of substitutions, deletions, and insertions, and $N$ is the number of reference units. Lower values indicate better performance.

\subsection{Network Configuration}
\label{sec:network_config}

\subsubsection{MVL Encoder}
The MVL encoder serves as the visual encoder of the proposed framework. It is initialized from AV-HuBERT Large~\cite{shi2022avhubert}. Following the AV-HuBERT configuration, the visual front-end uses a ResNet-18-based feature extractor~\cite{ma2021conformer,martinez2020lipreading}. The encoder consists of 24 Transformer blocks with 16 attention heads, a model dimension of 1024, and a feed-forward dimension of 4096.

\subsubsection{Audio Encoder and Audio-Visual Fusion Decoder}
The audio encoder is based on Whisper Large~\cite{radford2023robust}. Its front-end contains two 1D convolution layers with GELU activation. The audio-visual fusion decoder is built by modifying the Whisper decoder and inserting a visual cross-attention layer before self-attention in each decoder block. The decoder hidden states are used as queries, and the visual representations from the MVL encoder are used as keys and values. Each decoder block therefore contains visual cross-attention, self-attention, audio cross-attention, and a feed-forward network. In Whisper Large, both the encoder and decoder have 32 Transformer blocks with a model width of 1280 and 20 attention heads.

\subsubsection{Modality-Aware Module}
The Modality-Aware Module operates on the encoded audio and visual representations. It consists of a modality quality branch and a modality synchronization branch. The quality branch uses a temporal Conv1D layer followed by a two-layer MLP with ReLU and sigmoid activations to produce frame-level gating scores. The synchronization branch contains an audio projector and a visual projector, each implemented with two $1\times1$ Conv1D layers, with BatchNorm1D and ReLU between them, to project the two modalities into a shared synchrony embedding space.

\begin{table*}[t]
	\centering
	\caption{Comparisons with prior works on the LRS3 dataset. Results are reported on the original test set (Clean) and with babble noise at 0\,$\mathrm{dB}$ SNR (Noisy). We further evaluate performance under noisy conditions with different view angles ($5^\circ$, $15^\circ$) and different visual modality missing ratios ($0.1,0.3$). Noise dataset denotes the dataset used to generate babble noise for evaluation. $^{*}$ denotes large-scale pretrained LLM-based models evaluated without fine-tuning on LRS3. $^{\dagger}$ denotes the multi-view synthesized training data. `--' indicates that the corresponding result is not reported under this setting. The results of all methods are reproduced using the official open-source implementations.}
	\label{tab:exp1_overall}
	
	\begin{threeparttable}
		\setlength{\tabcolsep}{2.2pt}
		\renewcommand{\arraystretch}{1.05}
		
		\resizebox{0.99\textwidth}{!}{%
			\begin{tabular}{l c c c c c c c *{4}{c}}
				\toprule
				\multirow{2}{*}{\textbf{Model}} &
				\multirow{2}{*}{\textbf{Modality}} &
				\multirow{2}{*}{\begin{tabular}{@{}c@{}}\textbf{Total}\\\textbf{Params}\end{tabular}} &
				\multirow{2}{*}{\begin{tabular}{@{}c@{}}\textbf{Noise}\\\textbf{Dataset}\end{tabular}} &
				\multicolumn{2}{c}{\textbf{Training Data (hrs)}} &
				\multicolumn{2}{c}{\textbf{AVSR WER(\%)$\downarrow$}} &
				\multicolumn{4}{c}{\textbf{AVSR w.r.t Noisy WER(\%)$\downarrow$}} \\
				\cmidrule(lr){5-6}\cmidrule(lr){7-8}\cmidrule(lr){9-12}
				& & & &
				\textbf{Labeled} & \textbf{Unlabeled} &
				\textbf{Clean} & \textbf{Noisy} &
				\textbf{5$^\circ$} & \textbf{15$^\circ$} & \textbf{0.1} & \textbf{0.3} \\
				\midrule
				\noalign{\vskip -2pt}
				\rowcolor[gray]{0.88} \multicolumn{12}{c}{\textit{Supervised / Weakly Supervised methods}} \\
				V-CAFE~\cite{hong2022visual} & A+V & 109M & NoiseX & 433 & -- & 2.82 & 10.88 & 14.26 & 15.59 & 13.27 & 14.87 \\
				AV-RelScore~\cite{hong2023watch} & A+V & 136M & NoiseX & 433 & -- & 2.77 & 8.32 & 10.40 & 12.45 & 8.81 & 11.13 \\
				Auto-AVSR~\cite{ma2023auto} & A+V & 443M & NoiseX & 3,448 & -- & 0.90 & 2.00 & 4.63 & 6.86 & \textbf{4.07} & 6.11 \\
				Whisper-finetuned~\cite{rouditchenko2024whisper} & A & 1.6B & LRS3 & 433/1,759 & -- & 2.3/2.0 & 11.7/11.1 & -- & -- & -- & -- \\
				\midrule
				\noalign{\vskip -2pt}
				\rowcolor[gray]{0.88} \multicolumn{12}{c}{\textit{Self-supervised methods}} \\
				CMA~\cite{kim2024learning} & A+V & 505M & LRS3 & 433 & 1,759 & 1.50 & 4.40 & 7.94 & 11.93 & 6.33 & 9.35 \\
				AV-HuBERT~\cite{shi2022avhubert} & A+V & 477M & LRS3 & 433 & 1,759 & 1.40 & 5.80 & 9.60 & 14.08 & 7.29 & 11.14 \\
				Whisper-Flamingo~\cite{rouditchenko2024whisper} & A+V & 2.5B & LRS3 & 433/1,759 & 1,759 & 1.50/2.00 & 5.60/5.60 & 6.86/6.37 & 7.39/7.26 & 6.72/6.17 & 8.52/8.22 \\
				\midrule
				\noalign{\vskip -2pt}
				\rowcolor[gray]{0.88} \multicolumn{12}{c}{\textit{LLM-based methods}} \\
				Fun-ASR-Nano~\cite{an2025funasrtechnicalreport}$^{*}$ & A & 0.8B & LRS3 & millions & tens of millions & 2.96 & 26.18 & -- & -- & -- & -- \\
				FireRed-ASR~\cite{xu2025fireredasr}$^{*}$ & A & 1.1B & LRS3 & $\sim 8.1\times10^{4}$ & -- & 3.77 & 30.22 & -- & -- & -- & -- \\
				Qwen3-ASR~\cite{shi2026qwen3asr}$^{*}$ & A & 1.7B & LRS3 & -- & $\sim 4.0\times 10^{7}$ & 1.39 & 15.41 & -- & -- & -- & -- \\
				Llama-AVSR~\cite{Cappellazzo2025Large} & A+V & 8.7B & NoiseX & 1,759 & 1,759 & 0.77 & 4.00 & 6.00 & 9.48 & 4.73 & 6.75 \\
				MMS-Llama~\cite{yeo2025mms} & A+V & 3.2B & NoiseX & 433/1,759 & 1,759 & 0.90/0.72 & \textbf{2.4/1.9} & 4.90/3.95 & 6.93/6.19 & 4.79/4.53 & 6.17/5.90 \\
				\midrule
				\noalign{\vskip -2pt}
				\rowcolor[gray]{0.88} \multicolumn{12}{c}{\textit{Our method without multi-view data}} \\
				\textbf{M2S-AVSR} & A+V & 2.6B & LRS3 & 433/1,759 & 1,759 & \textbf{0.82/0.68} & 3.00/2.12 & 5.58/4.00 & 6.78/5.95 & 5.38/4.64 & 7.23/5.84 \\
				
				\midrule
				\noalign{\vskip -2pt}
				\rowcolor[gray]{0.88} \multicolumn{12}{c}{\textit{Comparison with multi-view data}} \\				
				Whisper-Flamingo~\cite{rouditchenko2024whisper} 
				& A+V & 2.5B & LRS3 
				& ms433/ms1,759$^{\dagger}$ & ms1,759$^{\dagger}$ 
				& 1.50/1.95 & 5.57/5.52 
				& 6.84/6.36 & 7.30/7.25 & 6.37/6.16 & 8.55/8.42 \\ 
				
				\textbf{M2S-AVSR} 
				& A+V & 2.6B & LRS3 
				& ms433/ms1,759$^{\dagger}$ & ms1,759$^{\dagger}$ 
				& \textbf{0.82/0.65} & 2.84/2.02 
				& \textbf{4.83/3.86} & \textbf{5.78/4.05} & 5.30/4.60 & \textbf{6.01/5.77} \\
				\bottomrule
			\end{tabular}%
		}
	\end{threeparttable}
\end{table*}

\subsection{Implementation Details}
\label{sec:implement_details}

For MVL encoder training, we initialize the model from the 5th-iteration AV-HuBERT Large~\cite{shi2022learning} and inherit its cluster pseudo-labels. The model is pretrained on ms433h and ms1759h, and then fine-tuned on ms30h or ms433h. We use a two-stage loss schedule with $\tau=0.07$ and $\alpha=0.6$, where $\lambda_{\mathrm{RDA}}$ is decayed from 0.3 to 0.1 in the last 70\% of training. The model is optimized with Adam (weight decay 0.01), using a learning rate of 0.002 with polynomial decay, and trained for 600k steps with 48k warm-up steps. For AV-HuBERT fine-tuning, we attach an attention-based sequence-to-sequence decoder with a token-level cross-entropy objective. For Whisper fine-tuning, all layers of Whisper Large are updated on audio-only data with additive noise. We use dynamic batching (max 72k frames), SpecAug~\cite{park2019specaugment}, and speed perturbation (0.9/1.0/1.1). The model is optimized with Adam using a learning rate of $1\mathrm{e}{-5}$ and 12k warm-up steps, and trained for 50k steps. For M2S-AVSR training, both encoders are frozen, while the visual cross-attention layers and the modality-aware module are trained from scratch. Following~\cite{chang2022on}, audio-visual modality dropout (0.5) is applied. The model is optimized with AdamW (weight decay 0.01) using a learning rate of $5\mathrm{e}{-5}$ and 5k warm-up steps, with dynamic batching (12k frames) and SpecAug. We set $T_w=2$ and train for 10k steps. All models are trained on two NVIDIA RTX PRO 6000 GPUs (96GB). During inference, beam search is used with a beam size of 25 and batch size of 4.

\section{Results and Discussions}
\label{sec:results}

\subsection{Overall Performance}
\label{sec:overall_performance}

As shown in Table~\ref{tab:exp1_overall}, we compare the proposed M2S-AVSR framework with previous approaches on the LRS3 dataset. Besides the clean condition, we also report results under babble noise at 0\,dB SNR. Robustness is further evaluated under noisy conditions using synthesized multi-view test samples with controlled viewpoint offsets ($5^\circ$, $15^\circ$) and partially masked lip regions with masking ratios of 0.1 and 0.3.

Overall, the proposed M2S-AVSR achieves competitive performance across all evaluation conditions, with particularly clear gains under noisy settings. It reduces the noisy WER from 5.80\% for AV-HuBERT and 4.40\% for CMA to 3.00\%, corresponding to relative reductions of 48.3\% and 31.8\%, respectively. Under the 1,759\,h setting, the noisy WER is further reduced to 2.12\%. Compared with LLM-based approaches, M2S-AVSR substantially outperforms Llama-AVSR under noisy conditions while achieving comparable performance to MMS-Llama with a smaller model size. Results under noisy conditions may not be strictly comparable due to differences in babble-noise generation protocols~\cite{rouditchenko2024whisper}. To further validate the robustness of the proposed framework under realistic conditions, additional evaluations are presented in Sections~\ref{sec:maf_results} and~\ref{sec:aishell8rs_benchmark}.

\begin{figure}[t]
    \centering
    \includegraphics[width=1.02\columnwidth]{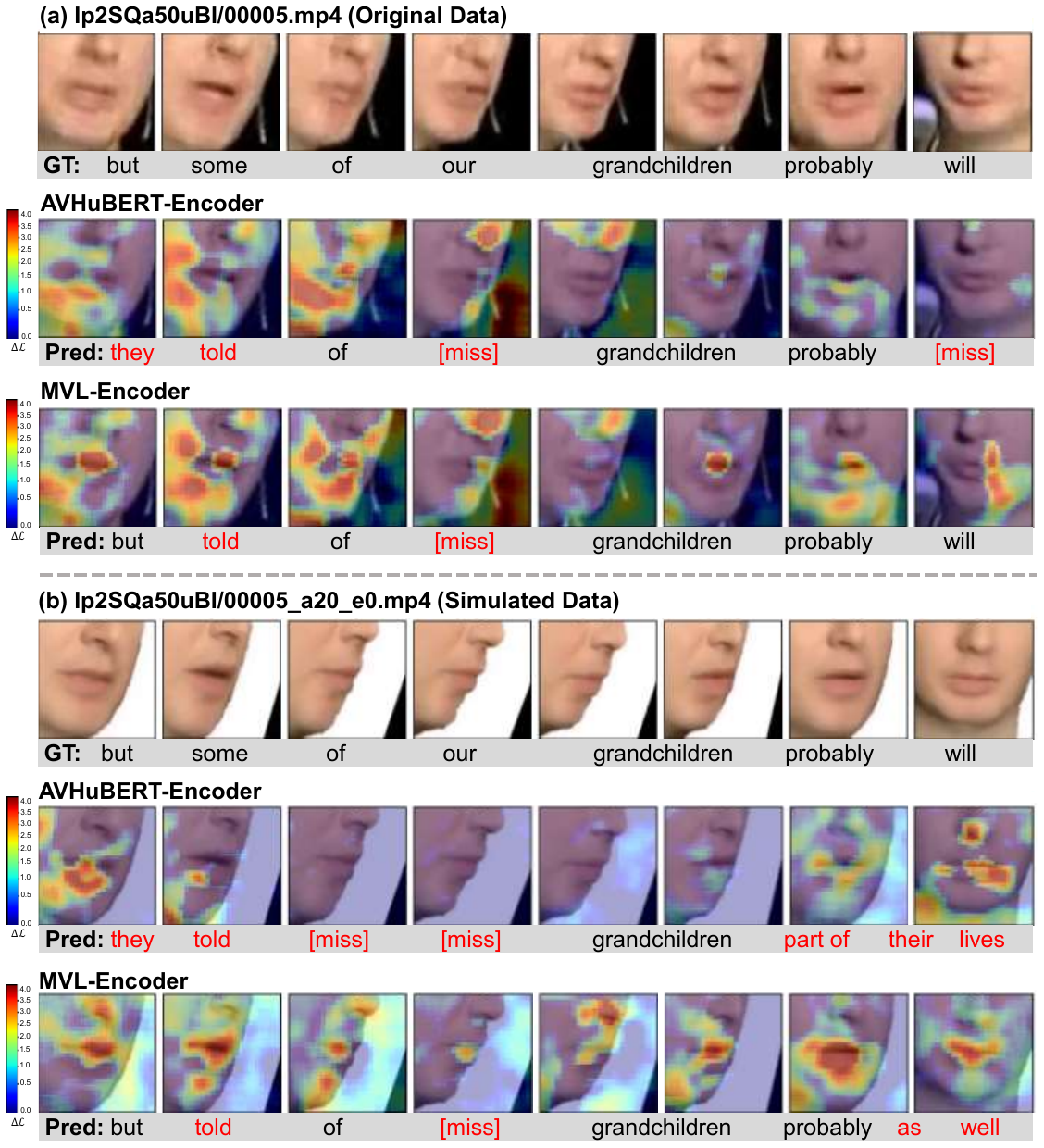}
        \caption{Occlusion-based lip sensitivity visualization for self-supervised multi-view representation learning on an LRS3 test utterance at SNR$=0$\,dB. (a) Original view. (b) Simulated view (yaw $=+20^\circ$), with identical audio input. Each row shows 8 uniformly sampled frames from a 60-frame clip. The heatmap color encodes the occlusion-induced loss change $\Delta\mathcal{L}$, where blue indicates lower values and red indicates higher values. \textbf{GT} and \textbf{Pred} denote the ground-truth and predicted transcripts, respectively. Red tokens in \textbf{Pred} indicate mismatches with \textbf{GT}, and \texttt{[miss]} denotes deletion.}
\label{fig_exper02_hotmap}
\vspace{-5pt}
\end{figure}

For robustness under viewpoint perturbation and visual degradation, M2S-AVSR consistently outperforms most existing AVSR methods. Compared with Whisper-Flamingo, it achieves relative WER reductions of 18.4\% and 18.0\% under the $5^\circ$ and $15^\circ$ viewpoint perturbation settings, respectively, and up to 15.5\% under visual masking conditions. It also achieves lower error rates than MMS-Llama under most robustness settings. To further evaluate the effectiveness of the proposed MVL encoder, we additionally compare M2S-AVSR and Whisper-Flamingo under the same multi-view training setting. With multi-view data, M2S-AVSR further reduces the noisy WER from 5.52\% to 2.02\% under the 1,759\,h setting, corresponding to a relative reduction of 63.4\%. It also achieves up to 29.4\% relative improvement under challenging visual conditions.

\subsection{Evaluation on Self-Supervised Multi-View Representation Learning}
\label{sec:eval_mvl}

We evaluate the proposed multi-view representation learning through visualization and comparison experiments.
For visualization, we adopt space--time occlusion sensitivity analysis~\cite{zeiler2014visualizing,zhou2015object} to examine the response of visual representations to viewpoint variation. Fig.~\ref{fig_exper02_hotmap} presents lip sensitivity maps on an LRS3 test utterance at 0\,dB SNR, comparing AV-HuBERT and MVL under both original and simulated views ($+20^\circ$ yaw). Under the original view, both models focus on lip regions, while the MVL encoder exhibits more concentrated responses. Under the simulated view, AV-HuBERT shows unstable responses with increased errors, whereas MVL maintains stable focus and yields more accurate predictions, indicating improved robustness to viewpoint variation. To further evaluate visual representation learning, we compare AV-HuBERT and MVL in VSR settings. Besides evaluations under different angular offsets and varying proportions of simulated multi-view data, we further evaluate the models on three LRS3 subsets grouped by face yaw angles and on the real multi-view OuluVS2 dataset. Across these evaluations, MVL consistently achieves lower WER than AV-HuBERT. Detailed quantitative results and analyses are provided in the supplementary material.


\subsection{Evaluation of Modality-Aware Fusion under Realistic Conditions}
\label{sec:maf_results}

We evaluate the proposed modality-aware module on the MISP2021-AVSR dataset, which contains real-world indoor recordings with noise, visual occlusion, and temporal instability. As shown in Table~\ref{tab:misp2021_fusion}, our method achieves a CER of 21.95\%, outperforming all previous systems. With Recognizer Output Voting Error Reduction (ROVER)~\cite{fiscus1997post}, the CER is further reduced to 18.82\%, corresponding to a 12.6\% relative reduction over the previous best result. These results demonstrate the effectiveness of modality-aware fusion under realistic conditions.

\begin{figure*}[t] 
    \centering 
    \includegraphics[width=\textwidth]{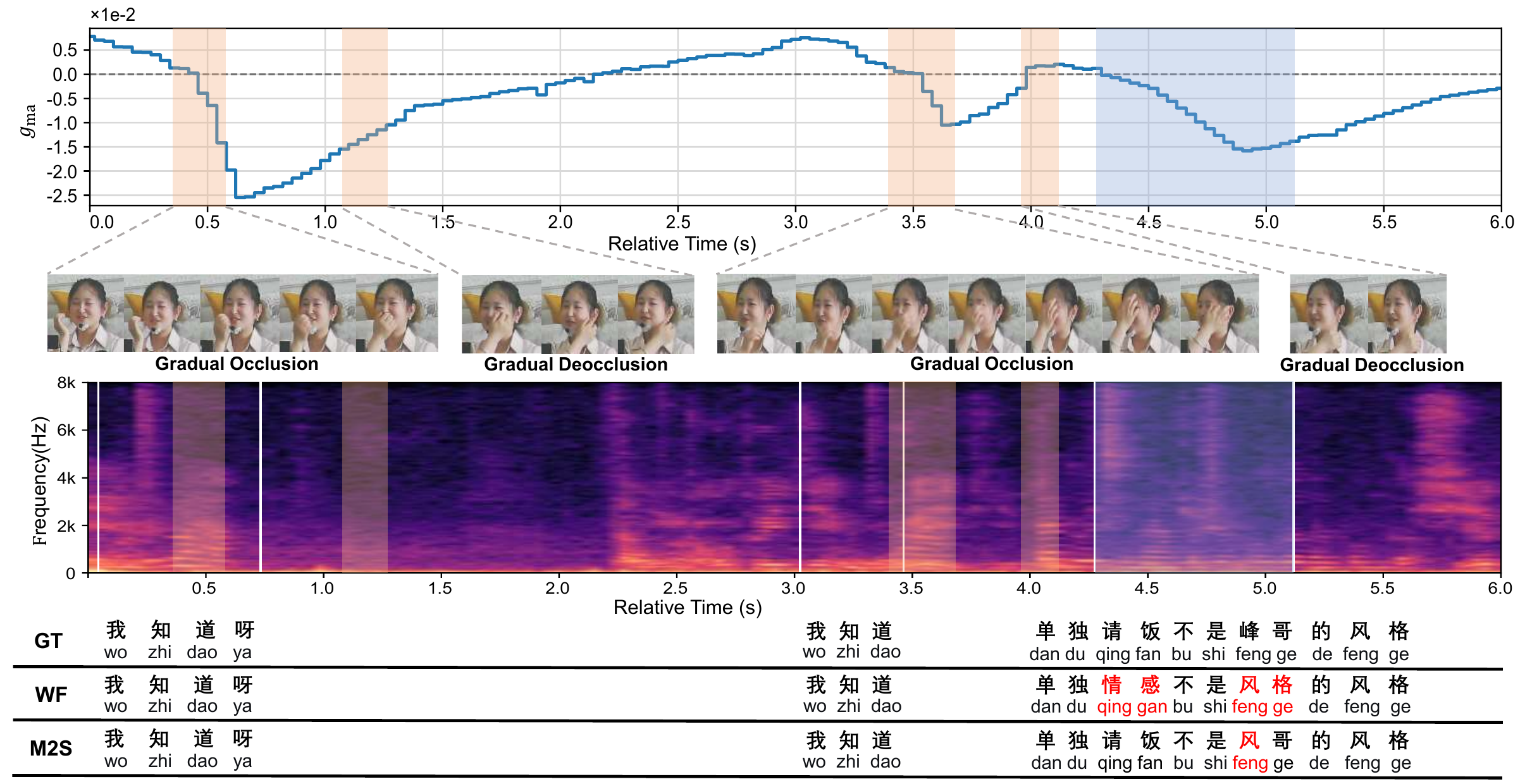}
    \caption{Visualization of modality-aware fusion on the MISP2021 dataset. 
    Test sample: R16\_S242243244245\_C02\_I1\_Far, Abs time: 1093--1099~s. 
    \textbf{Top:} Video-aligned modality-aware gating value $g_{\mathrm{ma}}$ over relative time (s).
    \textbf{Middle:} Facial snapshots illustrating gradual visual occlusion–deocclusion transitions.
    \textbf{Bottom:} Spectrogram and recognition outputs, including ground truth (GT), Whisper-Flamingo (WF), and the proposed M2S-AVSR (M2S). Red tokens mark transcript mismatches. Shaded regions indicate occlusion–deocclusion transitions (orange) and time spans aligned with recognition errors (blue).}
    \label{exper03_ma_visual} 
    \vspace{-10pt}
\end{figure*}

Fig.~\ref{exper03_ma_visual} illustrates the behavior of the modality-aware gate $g_{\mathrm{ma}}$ on a test segment. In regions with increasing visual occlusion, the gate value decreases, and it rises again when the lip region becomes visible, indicating adaptive regulation based on visual quality. In segments with weaker audio--visual consistency, the proposed module prevents over-reliance on a single modality, leading to more accurate recognition of ambiguous content. Overall, the proposed modality-aware fusion improves robustness by adaptively regulating cross-modal interaction under both visual corruption and cross-modal inconsistency.

\begin{table}[!t]
\caption{Comparison with prior systems on MISP2021 for evaluating modality-aware fusion under realistic conditions. ${\dagger}$ denotes initialized from Whisper pretraining. $^{\ddagger}$ The corresponding ROVER result reported in~\cite{dai2024study} is 21.53\%.}
\label{tab:misp2021_fusion}
\centering
\setlength{\tabcolsep}{2.2pt}
\renewcommand{\arraystretch}{0.95}
\footnotesize

\resizebox{\columnwidth}{!}{%
\begin{tabular}{l c c c c c}
\toprule
\multirow{2}{*}{\textbf{System}} &
\multirow{2}{*}{\textbf{Year}} &
\multicolumn{2}{c}{\textbf{Training Data (hrs)}} &
\multirow{2}{*}{\textbf{Backbone}} &
\multirow{2}{*}{\textbf{CER (\%)}$\downarrow$} \\
\cmidrule(lr){3-4}
& & \textbf{A} & \textbf{V} & & \\
\midrule

SJTU~\cite{wang2022sjtu}
& 2022
& 300
& LRW-1000~\cite{yang2019lrw}
& Conformer
& 34.02 \\

NIO~\cite{xu2022channel}
& 2022
& 3300
& LRW-1000
& Transformer
& 25.07 \\

USTC~\cite{meutzner2017improving}
& 2023
& 500
& w/o extra
& Conformer
& 24.58 \\

ModalBiasAVSR~\cite{dai2024study}
& 2024
& 1000
& w/o extra
& Conformer
& 22.13$^{\ddagger}$ \\

Whisper-Flamingo~\cite{rouditchenko2024whisper}
& 2025
& 600$^{\dagger}$
& LRS3+Vox2(En)
& Transformer
& 26.08 \\

\midrule

M2S-A
& \multirow{3}{*}{2026}
& 600$^{\dagger}$
& --
& \multirow{3}{*}{Transformer}
& 25.08 \\

M2S-V
&
& --
& LRS3+Vox2(En)
&
& 76.35 \\

\cellcolor{lightblue}\textbf{M2S-AV}
& \cellcolor{lightblue}
& \cellcolor{lightblue}600$^{\dagger}$
& \cellcolor{lightblue}LRS3+Vox2(En)
& \cellcolor{lightblue}
& \cellcolor{lightblue}\textbf{21.95} \\

\cellcolor{lightblue}\textbf{M2S-AV$_{\text{ROVER}}$}
& \cellcolor{lightblue}
& \cellcolor{lightblue}600$^{\dagger}$
& \cellcolor{lightblue}LRS3+Vox2(En)
& \cellcolor{lightblue}\textbf{}
& \cellcolor{lightblue}\textbf{18.82} \\

\bottomrule
\end{tabular}%
}

\vspace{-10pt}
\end{table}

\subsection{Benchmark Results on AISHELL8-RealScene}
\label{sec:aishell8rs_benchmark}

We establish a real-scene speech recognition benchmark on AISHELL8-RealScene, covering both large-scale pretrained speech models and recent LLM-based approaches.

\begin{table}[!t]
	\caption{Benchmark results on AISHELL8-RealScene (CER \%). Indoor and Outdoor results are averaged over three camera views (D0/D1/D2). Overall denotes the average CER across all scenes. $^{\star}$ denotes LLM-based models trained on large-scale data. Audio-only denotes the proposed M2S-AVSR trained and evaluated without visual modality.}
	\label{tab:aishell8rs_overall}
	\centering
	\setlength{\tabcolsep}{4pt}
	\renewcommand{\arraystretch}{1.05}
	\footnotesize
	\resizebox{\columnwidth}{!}{%
	\begin{tabular}{l c cc}
		\toprule
		\textbf{Model} & \textbf{Indoor}(5.28h) & \textbf{Outdoor}(6.37h) & \textbf{Overall}(11.65h) \\
		\midrule
		
		\multicolumn{4}{c}{\textbf{ASR Systems}} \\ \midrule
		Whisper-finetuned~\cite{rouditchenko2024whisper}
        & 29.49 & 42.01 & 35.75 \\
		Fun-ASR-Nano~\cite{an2025funasrtechnicalreport}$^{\star}$
        & 34.63 & 49.94 & 42.74 \\
		FireRed-ASR~\cite{xu2025fireredasr}$^{\star}$
        & \textbf{23.66} & 38.69 & \textbf{31.19} \\
		Qwen3-ASR~\cite{shi2026qwen3asr}$^{\star}$
        & 23.98 & 38.76 & 31.39 \\
		
		\midrule
		\multicolumn{4}{c}{\textbf{AVSR Systems}} \\ \midrule
		Whisper-Flamingo~\cite{rouditchenko2024whisper}
        & 28.30 & 41.64 & 34.97 \\
		LLaMA-AVSR~\cite{Cappellazzo2025Large}$^{\star}$
        & 27.43 & 40.82 & 34.13 \\
		MMS-LLaMA~\cite{yeo2025mms}$^{\star}$
        & 25.52 & 39.16 & 32.34 \\
		
		\midrule
		\multicolumn{4}{c}{\textbf{Proposed Method}} \\ \midrule
        M2S-AVSR (Audio-only) & 26.70 & 40.56 & 33.64 \\
		M2S-AVSR  & 25.35 & \textbf{37.47} & 31.41 \\
		
		\bottomrule        
	\end{tabular}
    }
    \vspace{-5pt}
\end{table}

\begin{table}[t]
	\centering
	\caption{\label{tab:ablation}
		Ablation study under the LRS3 433\,h training setting. Multi-View Data indicates whether multi-view training data are used. Noisy denotes babble noise added at SNR$=0$\,dB. View15$^\circ$ and Mask denote evaluation under the same noisy setting, with a 15$^\circ$ view angle and 30\% video frame masking, respectively.}

    \small
    \setlength{\tabcolsep}{2.2pt}
    \renewcommand{\arraystretch}{1.08}
	
	\resizebox{0.99\columnwidth}{!}{%
		\begin{tabular}{
				>{\raggedright\arraybackslash}m{2.55cm}
				ccccccccc
			}
			\toprule
			
			\multirow{2}{*}{\textbf{System}} &
			\multirow{2}{*}[-0.8ex]{\makecell[c]{\textbf{Multi-View}\\\textbf{Data}}} &
			\multicolumn{2}{c}{\textbf{MVL Encoder}} &
			\multicolumn{2}{c}{\textbf{Modality-Aware}} &
			\multicolumn{4}{c}{\textbf{WER (\%)$\downarrow$}} \\
			\cmidrule(lr){3-4}\cmidrule(lr){5-6}\cmidrule(lr){7-10}
			
			& & \textbf{$L_{\text{MVC}}$} & \textbf{$L_{\text{RDA}}$}
			& \makebox[1.1cm][c]{\textbf{$g_{\mathrm{q}}$}}
			& \makebox[0.50cm][c]{\textbf{$g_{\mathrm{s}}$}}
			& \textbf{Clean} & \textbf{Noisy} & \textbf{View15$^\circ$} & \textbf{Mask} \\
			\midrule
			
			\multirow{2}{*}{w/o MVL + MAF}
			& --         & -- & -- & -- & -- & 1.32 & 5.55 & 7.16 & 8.50 \\
			& \checkmark & -- & -- & -- & -- & 1.32 & 5.56 & 7.12 & 8.48 \\
			\midrule
			
			\multicolumn{10}{l}{\textbf{+ MVL Encoder}} \\
			\midrule
			
			w/ MVC
			& \checkmark & \checkmark & -- & -- & -- & 1.32 & 5.47 & 7.04 & 8.41 \\
			w/ RDA
			& \checkmark & -- & \checkmark & -- & -- & 1.32 & 5.60 & 6.94 & 8.45 \\
			w/ MVC + RDA
			& \checkmark & \checkmark & \checkmark & -- & -- & 1.32 & 5.43 & 6.05 & 8.42 \\
			\midrule
			
			\multicolumn{10}{l}{\textbf{+ Modality-Aware Fusion}} \\
			\midrule
			
			w/ QualityGate
			& \checkmark & \checkmark & \checkmark & \checkmark & -- & 1.10 & 2.93 & 6.14 & 6.37 \\
			w/ SynchronyGate
			& \checkmark & \checkmark & \checkmark & -- & \checkmark & 0.88 & 2.88 & 6.10 & 6.51 \\
			w/ Modality-Aware
			& -- & \checkmark & \checkmark & \checkmark & \checkmark
			& 0.82 & 2.90 & 6.78 & 7.23 \\
			w/ Modality-Aware
			& \checkmark & \checkmark & \checkmark & \checkmark & \checkmark
			& \textbf{0.82} & \textbf{2.84} & \textbf{5.78} & \textbf{6.01} \\
			\bottomrule
		\end{tabular}%
	}
	
	\vspace{-6pt}
	\end{table}

Table~\ref{tab:aishell8rs_overall} reports the results on indoor scenes, outdoor scenes, and the full evaluation set. All systems perform worse in real and noisy outdoor scenes, reflecting the stronger background noise and more severe visual interference in real-world environments. FireRed-ASR achieves the best overall CER of 31.19\%, benefiting from large-scale speech pretraining. However, audio-only modeling is less effective under more adverse outdoor conditions. On the outdoor subset, M2S-AVSR achieves the best CER of 37.47\%, corresponding to a 7.3\% relative reduction compared with the mean CER of Whisper-Flamingo and MMS-LLaMA. This indicates that incorporating visual information is beneficial for robust recognition in challenging real-scene environments.

We further analyze view-specific performance under different training-view settings in the supplementary material, where multi-view training is shown to improve robustness and consistency across camera views. Overall, these results highlight the importance of multi-view modeling and modality-aware fusion for real-world audio-visual speech recognition.

\subsection{Ablation Study}
\label{sec:ablation_study}

In this section, we conduct ablation experiments on LRS3 (433\,h). Starting from a baseline without MVL or modality-aware fusion, we first introduce the MVL encoder. Using either MVC or RDA alone yields limited gains, while combining both losses reduces the noisy WER from 5.55\% to 5.43\% and the 15$^\circ$ view WER from 7.16\% to 6.05\%, indicating more robust view-invariant representations. We then incorporate the modality-aware fusion module. Adding the quality-aware gate reduces the noisy WER to 2.93\%, and introducing the synchrony-aware gate further improves it to 2.88\%. When all components are combined, the model achieves 0.82\% WER on clean data and 2.84\% under noisy conditions, corresponding to a 48.8\% relative reduction compared with the system without MVL and modality-aware fusion. We also analyze the effect of multi-view data. When used alone, the improvement is limited, whereas combining multi-view data with the MVL encoder and modality-aware fusion consistently improves performance across all settings, showing that its benefit depends on appropriate modeling.

\section{Conclusions}
\label{sec:conclusions}

This paper presented M2S-AVSR, a robust audio-visual speech recognition framework for real-world environments. By combining multi-view representation learning with modality-aware fusion, the proposed method achieves robust audio-visual speech recognition under challenging conditions and attains state-of-the-art performance on MISP2021-AVSR. We also released AISHELL8-RealScene, a public multi-scenario, multi-view conversational audio-visual dataset recorded in real-world scenes, and established a real-scene AVSR benchmark covering large-scale pretrained ASR models and recent LLM-based systems. Experimental results validate the effectiveness of M2S-AVSR and demonstrate the value of AISHELL8-RealScene for future research.


\appendix[Additional Experimental Results]
\label{app:additional_results}

\section{Additional Experimental Results}

\subsection{Additional Results on Self-Supervised Multi-View Representation Learning}
\label{sec:sup_eval_mvl}

\begin{figure}[!t]
\centering
\includegraphics[width=1.00\columnwidth]{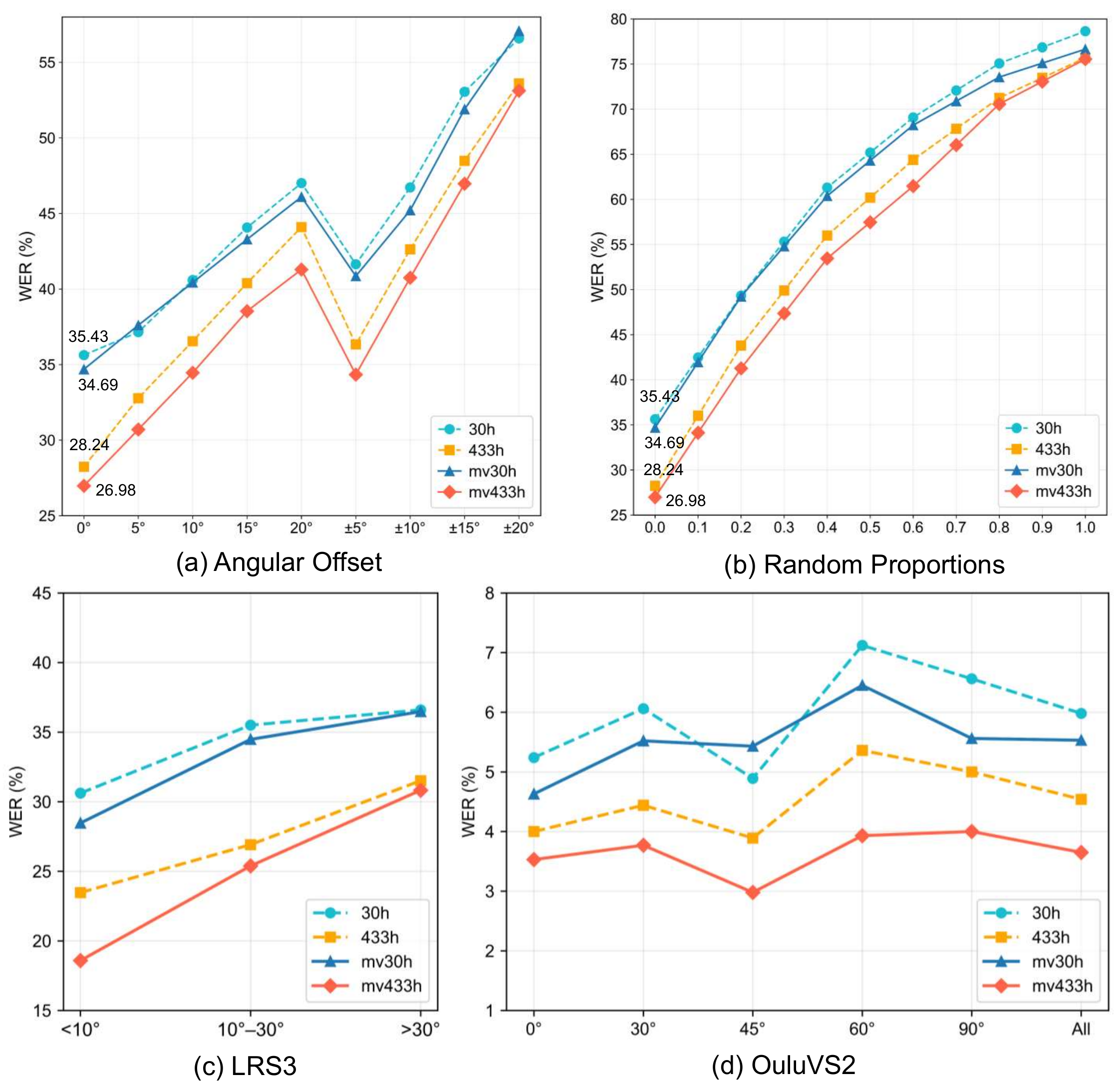}
\caption{
Evaluation on multi-view data (VSR WER (\%)).
``30h'' and ``433h'' denote AV-HuBERT Large pretrained on 1759h and fine-tuned on 30h and 433h, respectively.
``mv30h'' and ``mv433h'' denote AV-HuBERT with the MVL encoder, pretrained on ms1759h and fine-tuned on ms30h and ms433h, respectively. (a) Angular Offset: gradually introducing different horizontal viewing angles into the test set. (b) Random Proportions: evaluating varying proportions of simulated multi-view data. (c) LRS3: evaluation on three subsets of the LRS3 test set grouped by face yaw angles ($<10^\circ$, $10^\circ$--$30^\circ$, and $>30^\circ$). (d) OuluVS2: evaluation on five viewpoints ($0^\circ$, $30^\circ$, $45^\circ$, $60^\circ$, and $90^\circ$) and the overall set (All), following the protocol of \cite{Houjeung2017multiview}.}
\label{fig_exper02_curve}
\end{figure}

We provide additional quantitative results comparing the original AV-HuBERT encoder and the proposed MVL encoder under VSR settings. Fig.~\ref{fig_exper02_curve}(a) and (b) show the results on synthesized multi-view data. In Fig.~\ref{fig_exper02_curve}(a), MVL consistently achieves lower WER than the original AV-HuBERT models under different angular offsets. At $+20^\circ$, mv433h achieves a WER of 41.29\%, compared with 44.09\% for 433h, corresponding to a relative reduction of 6.37\%. Averaged over $0^\circ$ to $20^\circ$, the WER decreases from 36.41\% for 433h to 34.39\% for mv433h. Fig.~\ref{fig_exper02_curve}(b) shows a similar trend under different proportions of simulated multi-view data. For proportions from 0.2 to 0.6, mv433h achieves an average WER of 52.06\%, compared with 54.72\% for 433h, corresponding to a relative reduction of 5.0\%.

Fig.~\ref{fig_exper02_curve}(c) and (d) report real multi-view evaluations. For the LRS3 subsets grouped by face yaw angles in Fig.~\ref{fig_exper02_curve}(c), mv433h consistently achieves lower WERs than 433h, with relative reductions of up to 20.8\%. On the OuluVS2 dataset in Fig.~\ref{fig_exper02_curve}(d), mv433h achieves a WER of 3.65\% on the overall evaluation set, compared with 4.54\% for 433h, corresponding to a relative reduction of 19.6\%. Several models achieve their best performance around $45^\circ$, suggesting that moderate viewpoint offsets may provide complementary visual cues and lead to more discriminative visual speech representations.

\subsection{Additional Results on AISHELL8-RealScene}
\label{sec:aishell8rs_sup}

We further report view-specific results on AISHELL8-RealScene to analyze robustness across different camera views and training-view configurations.

\begin{table}[!t]
\caption{View-specific results on AISHELL8-RealScene (CER \%). Results are reported for each camera view (D0/D1/D2) and the average (Avg). Training Views denotes the views used during training: ``Single'' uses only the D1 view, while ``All'' uses all three views (D0/D1/D2).}
\label{tab:view_specific_sup}
\centering
\setlength{\tabcolsep}{4pt}
\renewcommand{\arraystretch}{1.2}
\footnotesize
\begin{tabular}{l c cccc}
\toprule
\multirow{2}{*}{\textbf{Model}} &
\multirow{2}{*}{\textbf{Training Views}} &
\multicolumn{4}{c}{\textbf{Overall}} \\
\cmidrule(lr){3-6}
& & D0 & D1 & D2 & Avg \\
\midrule
\multirow{2}{*}{Whisper-Flamingo~\cite{rouditchenko2024whisper}}
& Single, Front D1 & 35.05 & 34.48 & 34.75 & 34.76 \\
& All    & 35.12 & 34.76 & 35.03 & 34.97 \\
\midrule
\multirow{2}{*}{M2S-AVSR}
& Single, Front D1 & 32.04 & 31.73 & 31.99 & 31.92 \\
& All    & \textbf{31.41} & \textbf{31.41} & \textbf{31.41} & \textbf{31.41} \\
\bottomrule
\end{tabular}
\end{table}

Table~\ref{tab:view_specific_sup} reports the results for each camera view under different training-view settings. Under the single-view (front D1) training setting, M2S-AVSR achieves lower CER than Whisper-Flamingo across all test views (D0, D1, and D2), indicating improved generalization even when trained with limited visual viewpoints.

However, both models exhibit noticeable performance variation across different test views in the single-view training setting, suggesting unreliable robustness to viewpoint changes. In contrast, when trained with multi-view data, M2S-AVSR achieves similar performance across D0, D1, and D2, indicating strong robustness against view variability.

%
\bibliographystyle{IEEEtran}
\bibliography{main}


\end{document}